\documentclass[manuscript,screen,nonacm]{acmart}

\AtBeginDocument{%
  \providecommand\BibTeX{{%
    \normalfont B\kern-0.5em{\scshape i\kern-0.25em b}\kern-0.8em\TeX}}}

\usepackage{soul}
\usepackage[normalem]{ulem}

\usepackage{makecell}
\usepackage{multirow}

\newif\ifcomment

\ifcomment
\definecolor{moh_colour}{RGB}{255, 204, 204}
\newcommand{\moh}[1]{\sethlcolor{moh_colour}\hl{[Moh: #1]}}
\definecolor{ahm_colour}{RGB}{204, 204, 255}
\newcommand{\ahm}[1]{\sethlcolor{ahm_colour}\hl{[Ahm: #1]}}
\definecolor{ang_colour}{RGB}{50, 250, 100}
\newcommand{\ang}[1]{\sethlcolor{ang_colour}\hl{[Ang: #1]}}
\definecolor{pr_colour}{RGB}{255, 200, 0}
\newcommand{\pr}[1]{\sethlcolor{pr_colour}\hl{[Pending Review: #1]}}
\newcommand{\tempremoved}[1]{}
\else
\newcommand{\ahm}[1]{}
\newcommand{\moh}[1]{}
\newcommand{\ang}[1]{}
\newcommand{\pr}[1]{}
\newcommand{\tempremoved}[1]{}

\fi
\renewcommand{\comment}[1]{}

\definecolor{revision_colour}{RGB}{0,255,255}
\newcommand{\revision}[1]{#1}

\definecolor{ommission_colour}{RGB}{255,0,0}
\newcommand{\ommission}[1]{}

\usepackage{subcaption}
\begin{document}

\title[PQA: Product Quantization DNN Hardware Accelerator]{PQA: Exploring the Potential of Product Quantization in DNN Hardware Acceleration}

\author{Ahmed F. AbouElhamayed}
\authornote{Both authors contributed equally to this research.}
\email{afa55@cornell.edu}
\orcid{0000-0001-6381-2936}
\author{Angela Cui}
\authornotemark[1]
\email{ayc62@cornell.edu}
\affiliation{%
  \institution{Cornell University}
  \state{New York}
  \country{USA}
}

\author{Javier Fernandez-Marques}
\affiliation{
 \institution{Flower Labs}
 \country{UK}}
\email{javermarq@gmail.com}

\author{Nicholas D. Lane}
\affiliation{%
 \institution{University of Cambridge}
 \country{UK}}

\author{Mohamed S. Abdelfattah}
\affiliation{%
  \institution{Cornell University}
  \state{New York}
  \country{USA}}
\email{mohamed@cornell.edu}
\orcid{0000-0002-4568-8932}

\renewcommand{\shortauthors}{AbouElhamayed, Cui, Fernandez-Marques, Lane, and Abdelfattah}

\begin{abstract}
Conventional multiply-accumulate (MAC) operations have long dominated computation time for deep neural networks (DNNs), espcially convolutional neural networks (CNNs). 
Recently, product quantization (PQ) has been applied to these workloads, replacing MACs with memory lookups to pre-computed dot products.
To better understand the efficiency tradeoffs of product-quantized DNNs (PQ-DNNs), we create a custom hardware accelerator to parallelize and accelerate nearest-neighbor search and dot-product lookups. 
Additionally, we perform an empirical study to investigate the efficiency--accuracy tradeoffs of different PQ parameterizations and training methods.
We identify PQ configurations that improve performance-per-area for ResNet20 by up to 3.1$\times$, even when compared to a highly optimized conventional DNN accelerator, with similar improvements on two additional compact DNNs.
When comparing to recent PQ solutions, we outperform prior work by $4\times$ in terms of performance-per-area with a 0.6\% accuracy degradation.
Finally, we reduce the bitwidth of PQ operations to investigate the impact on both hardware efficiency and accuracy. 
With only 2--6-bit precision on three compact DNNs, we were able to maintain DNN accuracy eliminating the need for DSPs.
\end{abstract}



\maketitle

\section{Introduction}
\label{sec:introduction}

Deep Neural Networks (DNNs) have become an essential computing technique and is finding its way to many computing form factors.
Edge computing is especially challenging because of the constrained computing environment and limited power budget.
To address this, DNN model optimization has taken many forms, including pruning~\cite{MLSYS2020_d2ddea18}, quantization~\cite{xnor,Jacob_2018, Wang_2019_CVPR}, lightweight architectural designs~\cite{mobilenetV2,tan2019efficientnet,pan2022edgevits}, and faster algorithms~\cite{Lavin_2016,strassenTschannen,Chen_2020_CVPR}. 
Each of these techniques offers different trade-offs in terms of inference acceleration, model compression, hardware acceleration suitability, and accuracy degradation. 
Some methods focus solely on decreasing the number of computations while not affecting model size (e.g. Winograd~\cite{Lavin_2016}).
Other methods such as linear quantization and pruning tend to generally decrease both computation and memory footprint.
Alternatively, methods such as some forms of non-linear quantization~\cite{Stock2020And} only decrease model size, and leave the number of computations unaffected.
On this spectrum of compute and memory tradeoffs, a new compression methodology, product quantization (PQ), sits on one extreme.
Specifically, PQ can eliminate \textit{all} multiplications from the matrix multiplication operation~\cite{pmlr-v139-blalock21a}, making it an interesting new compression method for approximating DNNs worth further investigation.
Our work thoroughly investigates this emerging methodology and determines its practicality and hardware acceleration potential.

PQ emerges from the research area of information retrieval.
Specifically, approximate nearest neighbours for information retrieval involves extracting a compact representation of high-dimensional features, for sample images~\cite{Yu_2018_ECCV, Jang_2020_CVPR} in order to fetch similar ones from a data source or database. 
These input features can be encoded using PQ~\cite{5432202}. 
More recently, PQ has been repurposed for DNN inference acceleration---specifically, by accelerating matrix multiplication~\cite{pmlr-v139-blalock21a}---by encoding layer inputs into a set of \textit{learnable} \textit{prototypes}\revision{ that replace inputs during inference according to the closest prototype. The set of prototypes is knowns as the prototype table}. 
This allows a pre-computed dot product between inputs and parameters to be fetched from a lookup table instead of performing the conventional multiply-accumulate operations. 
However, existing works offer a limited evaluation of this new compression paradigm.
For example, by exclusively applying PQ to the final layer of the DNN~\cite{pmlr-v139-blalock21a}, through extremely simple networks~\cite{pq_all_you_need}, or by ignoring the efficiency implications associated with PQ.
This is the case of PECAN~\cite{pecan}, which uses short codes to lessen the accuracy degradation that PQ introduces at the cost of a 20$\times$ increase in memory footprint, resulting in an overall slowdown in DNN execution despite the elimination of multiply-accumulate operations. 

Product quantization accelerates DNN inference by replacing convolutions (and in general any type of layer doing matrix-matrix multiplication) by a series of memory look-ups of pre-computed partial dot-products. 
Compute speedup is therefore achieved by reducing the computational footprint of compute intensive layers (for example, convolutions), sometimes in exchange of a higher memory footprint (depending on PQ parameters as we show in this work). 
As shown in previous works~\cite{pmlr-v139-blalock21a, pecan}, PQ has a big potential but it remains unclear how to effectively use this technique to accelerate DNNs without incurring severe accuracy degradation or a large memory footprint. 
Previous work (PECAN~\cite{pecan}) focused on demonstrating the potential of PQ but disregarded its efficiency trade-offs.
For example, prior work did not evaluate whether PQ can actually result in compute speedups, and the resulting model sizes were many times larger than the original (non-PQ) model~\cite{pecan}. 
As we demonstrate in our work, accounting for FLOPs only is not a guarantee for speedup. 
This is because PQ replaces compute with memory accesses, and these are not captured when reporting FLOPs. 
Our work is the first to provide a more holistic efficiency analysis, which is fundamental for future work proposing alternative PQ implementations, for example, with new encoding functions, lightweight distance metrics, or new training methodologies.

Our work also proposes the first custom hardware architecture for DNN acceleration with product quantization---the Product Quantization Accelerator (PQA).
We quickly realized that running PQ-DNNs on commodity CPUs or GPUs does not adequately reflect its hardware speedup potential, simply because these hardware options operate in very different ways to PQ-DNNs.
Instead, we create a custom PQA on Intel Agilex FPGAs, taking advantage of a custom on-chip memory hierarchy and leveraging parallelism in PQ processing during nearest-neighbour computation, partial product look-up, and accumulation.
Furthermore, we explore the use of low numerical bitwidth in these PQ operations. 
This fundamentally differs from traditional DNN quantization work that uses low bitwidths to approximate matrix multiplication \revision{where quantization here is used in components used for distance calculation and in stored results in $LUT_{PQ}$ unlocking some new options explored in Section}~\ref{sec:pq_quantization}.

Motivated by the potential of PQ for inference acceleration, our work performs a holistic study of PQ in DNNs by exploring both its algorithmic efficiency, training dynamics, and hardware implementation.
More concretely, our contributions are enumerated below:
\begin{enumerate}
    \item Present the first Product Quantization Accelerator (PQA) for lightweight CNNs, demonstrating that unlike GPUs and CPUs, custom hardware can indeed accelerate PQ by up to 3.1$\times$ compared to conventional DNNs.
    \item Evaluate opportunities for low numerical bitwidth, with hardware configurations that maintain accuracy with only 2-bit distance calculation operations on a Keyword Spotting task.
    \item Evaluate PQ comprehensively for three CNNs (ResNet-20$_{CIFAR10}$, DW$_{EMNIST}$, MicroNet$_{KWS}$) to identify configurations that achieve 4--6$\times$ higher throughput compared to the most recent PQ literature, with minor accuracy loss ($\sim$0.6--0.9\%).
    \item Provide a systematic study on the parameterization and training of PQ and its impact on compute and memory footprint and encoding degradation, including the proposal of a \revision{"corrector"} DNN to improve PQ-DNN accuracy.
\end{enumerate}

\section{Product Quantization for DNNs}
\label{sec:bkg}

\begin{figure*}[t]
    \centering
    {
    \includegraphics[width=1\linewidth, trim=1.4cm 0 1.3cm 0]{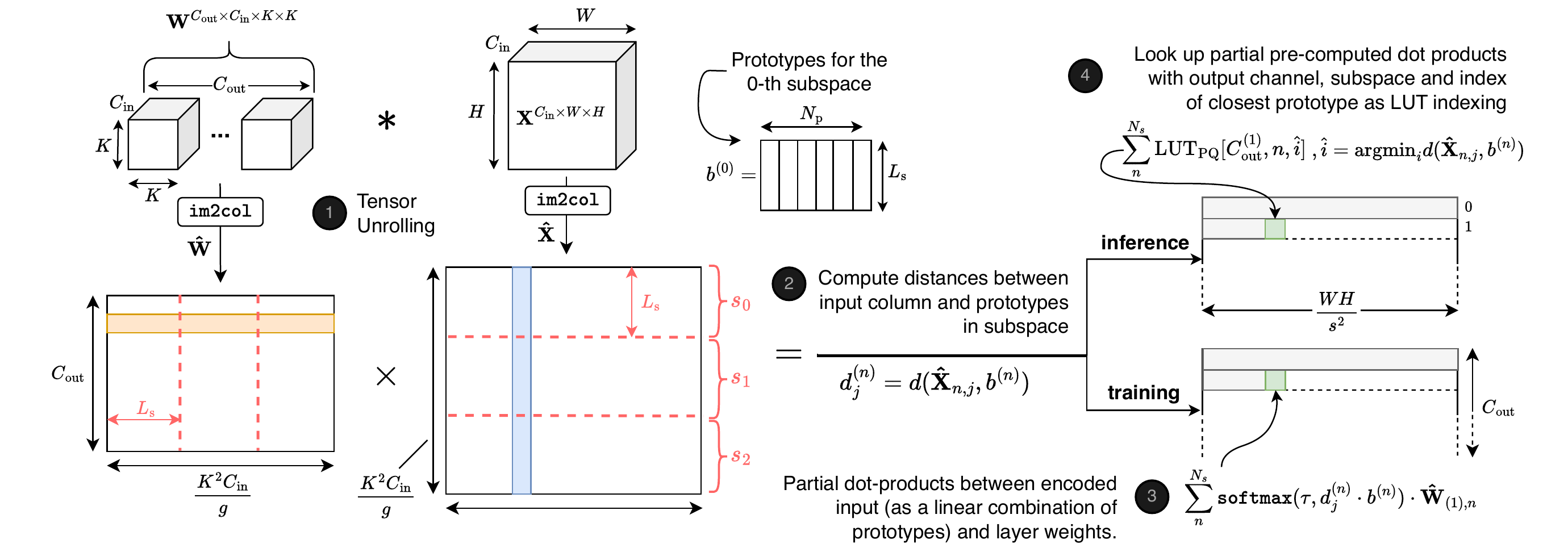}
    }
    \vspace{-3mm}
    \caption{Transforming a convolutional layer into its PQ equivalent. First, both input and weights tensors need to be unrolled, resulting in $\mathbf{\hat{X}}$ and $\mathbf{\hat{W}}$. The input matrix $\mathbf{\hat{X}}$ is subdivided into $N_s$ subspaces (three in this diagram), and the sub-columns in each one will be encoded using their respective bank of of prototypes $\mathbf{B}_{l}\!=\! [\mathbf{b}^{(0)}, \mathbf{b}^{(1)}, \mathbf{b}^{(2)}]$. Given a distance metric $d(\cdot)$, the input is encoded in a soft manner during training but replaced with a hard one-hot encoding during inference. For deployment, all the pre-computed dot products are stored in $\text{LUT}_{\text{PQ}}$.}
    \vspace{-3mm}
    \label{fig:tinypq_diagram}
\end{figure*}

This section explains the process of training and inference for Product-Quantized DNNs to eliminate multiplications from the dot product operations prevalent in DNNs. 
As illustrated in Figure~\ref{fig:tinypq_diagram}, we explain how convolutional or fully-connected layers can be approximated with PQ during training, and how to deploy a PQ-DNN after it's trained.

\subsection{Product Quantization Fundamentals}

Almost all layers found in modern ML architectures, including those in CNNs~\cite{He_2016, mobilenetV2, tan2019efficientnet} or transformers~\cite{NIPS2017_3f5ee243,mehta2022mobilevit}, can be expressed as a matrix-matrix multiplication. 
This is evident for standard fully connected layers where an input $\mathbf{X}\!\in\!\mathbb{R}^{A\times C_{\text{in}}}$ is transformed by weights $\mathbf{W}\!\in\!\mathbb{R}^{C_{\text{out}}\times A}$ to obtain output $\mathbf{Y}\!\in\! \mathbb{R}^{A\times C_{\text{out}}}$, where $C_{\text{in}}$ and $C_{\text{out}}$ are the number of input and output features or, more generically, channels. In the case of convolutions, both input $\mathbf{X}\!\in\! \mathbb{R}^{C_{\text{in}}\times W \times H}$ and weights $\mathbf{W}\!\in\! \mathbb{R}^{C_{\text{out}}\times C_{\text{in}}\times K_{\text{w}} \times K_{\text{h}}}$ tensors first need to be unrolled. This unrolling can be done following the {\tt im2col} algorithm~\cite{Jia:EECS-2014-93, caffe}, resulting in unrolled matrices of size $(K_{w}K_{h}C_{\text{in}}/g) \times (WH/s^2)$ and $C_{\text{out}} \times (K_{\text{w}}K_{\text{h}}C_{\text{in}}/g)$ for input and weights tensors respectively with $s$ and $g$ representing stride and groups respectively.
This is illustrated in Figure~\ref{fig:tinypq_diagram} (left).

We now introduce PQ-specific terms and assume a matrix of weights $\mathbf{W}\!\in\! \mathbb{R}^{C_{\text{out}}\times A}$ and of inputs $\mathbf{X}\!\in\! \mathbb{R}^{A\times C_{\text{in}}}$ exist, whether they are from a linear layer or are the result of a tensor unrolling. 
With PQ, we aim to aggressively quantize the input matrix on-the-fly so that the multiplication with $\mathbf{W}$ can be reduced to a series of lookups to a table of pre-computed dot-products of size $\text{LUT}_{\text{PQ}}\!\in\! \mathbb{R}^{C_{\text{out}}\times N_{\text{s}} \times N_{\text{p}}}$. 
The input $\mathbf{X}$ can be split into $N_{\text{s}}$ disjoint groups along the rows dimension. We call each of these groups \textit{subspaces} as shown in Figure~\ref{fig:tinypq_diagram}. 
Each subspace gets assigned a bank of \textit{prototypes} $\mathbf{b}^{(n)}$ of size $L_{\text{s}}\times N_{\text{p}}$, where $N_{\text{p}}$ and $L_{\text{s}}$ represent the number and the length of the prototypes respectively. %
This list of prototypes banks $\mathbf{B}_{l}\!=\![\mathbf{b}^{(0)}, \mathbf{b}^{(1)}, ..., \mathbf{b}^{(N_{\text{s}}-1)}]$ are learnable parameters of PQ layers, with subscript $l$ denoting the layer index. 
PQ involves getting the index of the closest prototype via distance $d(\cdot)$ to each sub-column of the input matrix (\textit{i.e.} vectors of length $L_{\text{s}}$). 
In addition, all dot products between the weights and the prototypes are precomputed and stored in a look-up table $\text{LUT}_{\text{PQ}}$.
Therefore, during inference, a dot product is looked up from $\text{LUT}_{\text{PQ}}$ depending on the closest prototype matched by an input vector, eliminating multiplication operations altogether.

Training PQ layers involves learning two sets of parameters: the standard layer weights $\mathbf{W}_{l}$ and prototypes $\mathbf{B}_{l}$. 
Then, prior to deployment, the table of pre-computed dot products $\text{LUT}_{\text{PQ}}$ is obtained by performing a Cartesian product between $\mathbf{W}_{l}$ and $\mathbf{B}_{l}$. 
Only $\text{LUT}_{\text{PQ}}$ and $\mathbf{B}_{l}$ are deployed to perform perform inference, with no need for the actual model parameters $\mathbf{W}_{l}$.
This is because $\text{LUT}_{\text{PQ}}$ explicitly contains all model parameters and their dot products with the input prototypes $\mathbf{B}_{l}$.
Figure~\ref{fig:tinypq_diagram} (right) illustrates training and inference for PQ-DNNs, explained further below.

\subsection{Choosing a distance metric}
For both training and inference, a distance $d(\cdot)$ needs to be defined to encode the input using the layer prototypes $\mathbf{B}_{l}$. 
A reasonable choice would be the Euclidean distance. 
For PQ, the relative ranking of distances between input column $\mathbf{\hat{X}}_{n,j}$ and the prototypes of that subspace $\mathbf{b}^{(n)}$ is far more important than the actual distance value. 
This therefore leaves room for more lightweight distance metrics such as the $L_1$ Manhattan distance as well.
Recent work~\cite{pmlr-v139-blalock21a} has also proposed using locality-sensitive hashing to match inputs with prototypes thereby performing comparisons between inputs and prototypes more quickly but placing further constraints on the nature of the learned prototypes $\mathbf{B}_{l}$.
In our case, we explore Euclidean distance (L2 norm). 
While this introduces multiplication to compute the squared distance, performance is still dominated by memory access. 
We also explore Manhattan distance (L1 norm) but that often led to higher accuracy degradation.

\subsection{Input encoding} 
During training the encoding of the input is done explicitly, \textit{i.e.}, the unrolled input $\mathbf{\hat{X}}$ is actually transformed into a new matrix of the same dimensions that is later multiplied with the unrolled layer weights $\mathbf{\hat{W}}$. 
Given the $j$-th column of $\mathbf{\hat{X}}$ along the $n$-th subspace, $\mathbf{\hat{X}}_{n,j}$, the distances $\mathbf{d}^{(n)}_j = d(\mathbf{\hat{X}}_{n,j}, \mathbf{b}^{(n)})$ to each of the prototypes are computed. 
The encoded portion of the input $\mathbf{\hat{X}^{\text{enc}}}_{n,j}$ is obtained by a weighed linear combination of $\mathbf{b}^{(n)}$ with normalized distances $\phi(\tau,\mathbf{d}^{(n)}_j)$

\begin{equation}
    \mathbf{\hat{X}^{\text{enc}}}_{n,j} = \phi\left(\tau,\mathbf{d}^{(n)}_j\right) \mathbf{b}^{(n)} = \sum^{N_{\text{p}}}_p  \frac{\mathbf{b}^{\text{(n)}}_p\text{exp}(d^{(n,p)}_j/\tau)}{\sum^{N_{\text{p}}}_k\text{exp}(d^{(n,k)}_j/\tau)} 
\label{eq:encoding}
\end{equation}

\noindent where $\mathbf{b}^{\text{(n)}}_p$ stands for the $p$-th prototype in the $n$-th subspace, $d^{(n,p)}_j$ is the distance to the p-th prototype, and $\tau$ is a temperature factor. 
As $\tau\rightarrow0$, the temperatured-softmax $\phi(\tau, \cdot)$ outputs a sharper distribution over $\mathbf{d}^{(n)}_j$, transitioning in this way from \textit{soft} linear combination of prototypes \revision{where all prototypes are considered with different weights} into a \textit{hard}, one-hot encoding\revision{ where only the prototype closest to the input is considered}. 
During inference PQ is implemented with one-hot assignments to match a single prototype to each input vector.

\subsection{Constructing $\text{LUT}_{\text{PQ}}$} 
At inference, $\mathbf{\hat{X}^{\text{enc}}}_{n,j}$ is not materialised since only the index of the closest prototype is needed to retrieve the partial dot product from $\text{LUT}_{\text{PQ}}$. After training, the unrolled weights $\mathbf{\hat{W}}$ are partitioned into subspaces along the column dimension (as shown in Figure~\ref{fig:tinypq_diagram}). Then, the dot product between a $1\times L_{\text{s}}$ sub-row in $\mathbf{\hat{W}}$ and a prototype in such subspace, corresponds to one entry in $\text{LUT}_{\text{PQ}}$. 
Repeating this for all $N_{s}$, $C_{\text{out}}$ and $N_{\text{p}}$ completes the table.
This allows us to fetch a precomputed dot product between any prototype $\mathbf{B}_{l}$ and its corresponding layer weights.

\subsection{Exploring Low Numerical Bitwidth for PQ-DNNs}
\label{sec:pq_quantization}

Although PQ is fundamentally a vector quantization method, it offers scope for applying \textit{additional} quantization to its constituent operations.
Specifically, lower bitwidth can be used within the prototype table and the $LUT_{PQ}$, thereby offering significant efficiency and area advantages.
We apply post-training asymmetric linear quantization for both of these operations. 
For the prototype table, we abstain from dequantizing the stored quantized values. 
Instead, the incoming input is quantized using identical parameters, and the distance calculation is performed within the quantized domain. As we see in the results, this approach yields satisfactory results. In contrast, for the $LUT_{PQ}$, we employ dequantization of stored quantized values to perform wide accumulation in 16-bit precision.
Determining the scale and offset for quantization entails various possibilities, each bringing its unique cost and benefit trade-off. 
The most cost-effective option is to have a single scale and offset value for the entire model. 
Per-layer and per-channel quantization parameters are also prevalent in conventional quantization to improve accuracy.
Furthermore, we introduce a novel \textit{per-subspace} quantization parameters---uniquely-suited to PQ-based matrix multiplication, and as we show in the results, achieve the highest accuracy.

\subsection{The Potential for Compute Speedups on CPU, GPU, and PQA}
\label{sec:pq_speedup_on_different_hardware_platforms}

\begin{figure}[t]
    \includegraphics[width=\linewidth]{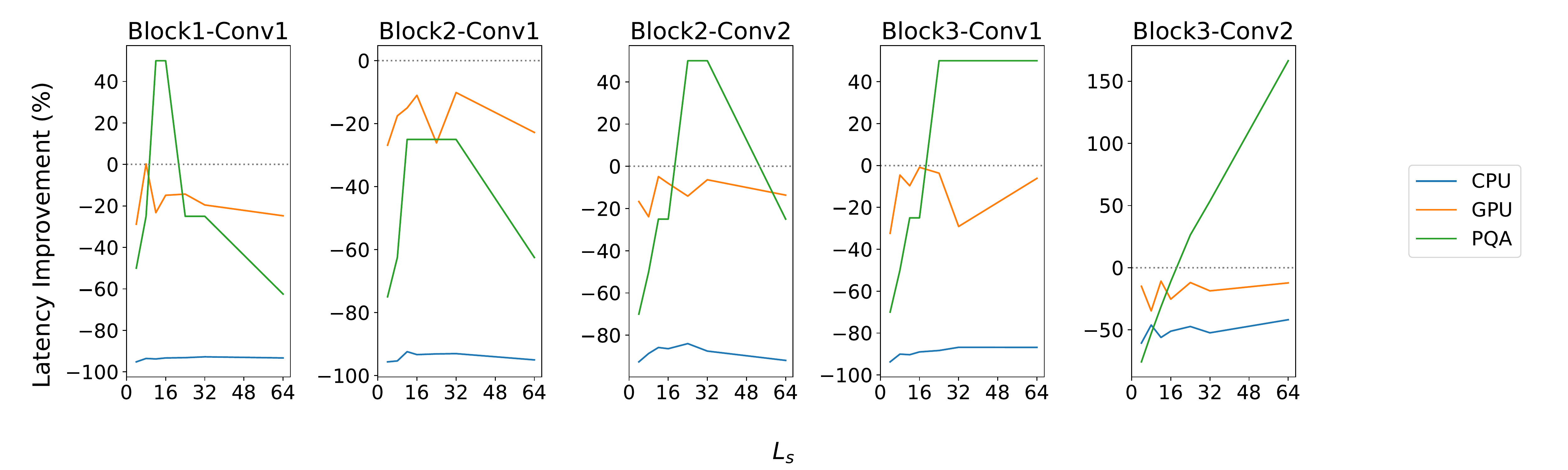}
    \vspace{-6mm}
    \caption{Speedup of PQ vs conventional convolutions on different hardware when run on all unique layers of ResNet20. \revision{CPU is 13th Gen Intel(R) Core(TM) i9-13900K and GPU is NVIDIA GeForce RTX 4090.}}
    \label{fig:speedup_on_different_hardware}
\end{figure}

Commodity hardware such as CPUs and GPUs are inherently unsuitable for PQ-DNNs but are increasingly better suited to conventional DNNs, making a comparison on those platforms somewhat skewed. 
A key reason for creating PQA is to be able to fairly assess the efficiency of PQ-DNNs compared to conventional DNNs. PQA has the potential to attain high efficiency because of the high levels of parallelism and memory banking that is possible as described in Section~\ref{sec:hw_model}.
To quantify this, we measure the speedup of executing PQ layers of ResNet20~\cite{pecan} on a CPU, GPU, and PQA. 
Our baseline custom hardware is DLA~\cite{dla_new} for native convolutions.
Figure~\ref{fig:speedup_on_different_hardware} shows the speedup percentage for the unique layers of the network, obtained by running PQ and conventional convolution on different hardware and comparing the latencies.
We sweep different values of $L_\text{s}$ while $N_\text{p}$ is kept at 16. 
It is clear that there is a consistent slowdown when running PQ on a CPU or a GPU as there is no speedup for different values of $L_s$ and the same trend appears when trying different values of $N_\text{p}$. 
Measuring the effect of PQ on commodity hardware like CPUs or GPUs is therefore a poor way of assessing its acceleration potential.
However, using a custom hardware that is specifically designed for PQ achieves improvements in latency for \textit{certain} configurations over DLA~\cite{aydonat2017opencl,dla_new}.
While this is not the case in all layers, the speedup reaches up to 150$\%$ in the last layer of the network at a large value of $L_s$.

\section{Product Quantization Accelerator (PQA) Architecture}
\label{sec:hw_model}

In this section we detail the design of a custom PQ inference Accelerator (PQA) on an FPGA.
We also benchmark PQ-DNNs on CPUs and GPUs, and we argue with empirical results that these devices are not suitable for assessing the efficiency potential of PQ.
This motivates the design of PQA which we show to indeed result in faster DNN execution for some PQ parameters.

\begin{figure*}[t]
    \centering
    {
    \includegraphics[width=0.85\linewidth, trim=.5cm .1cm 0 1cm]{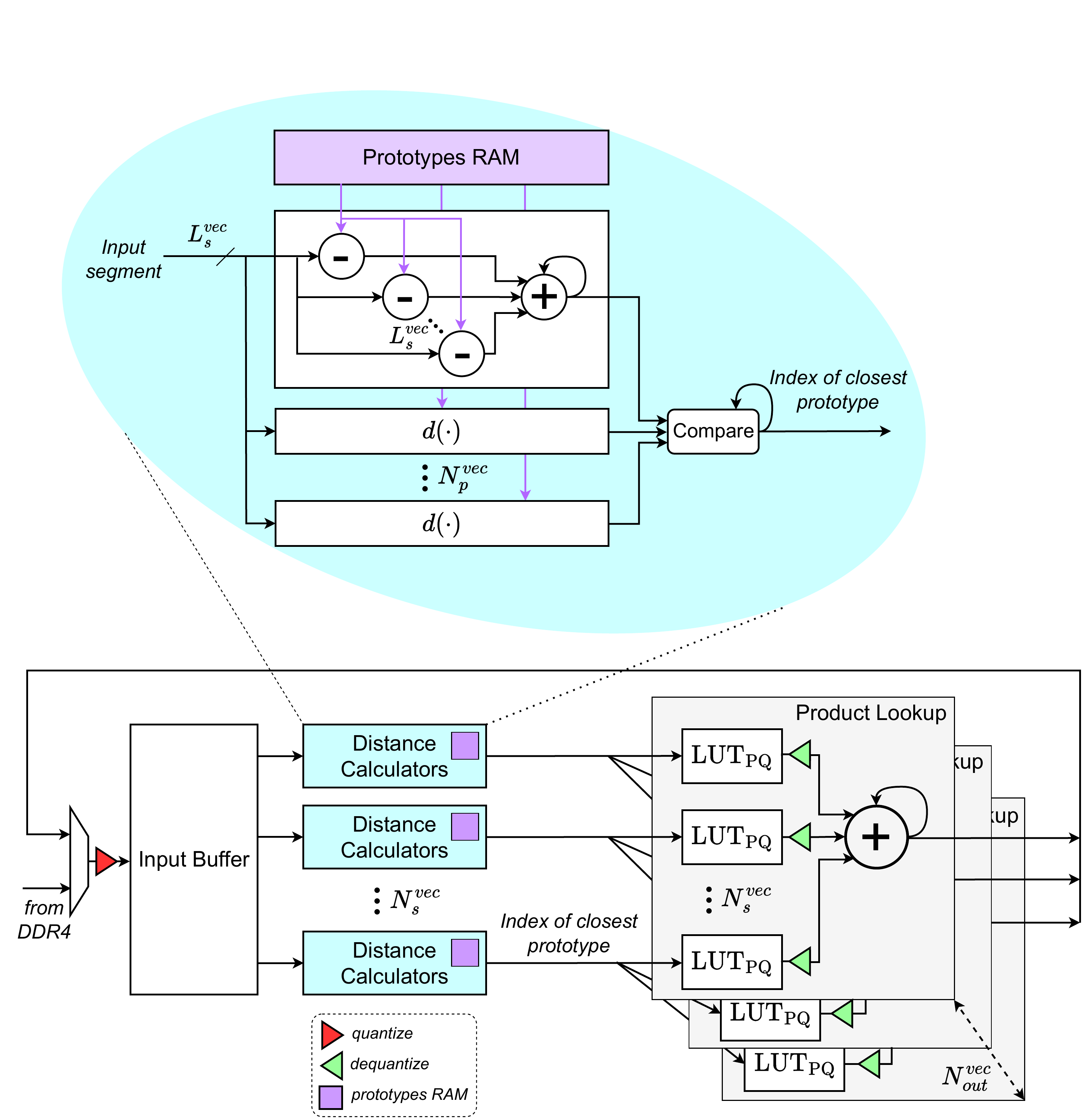}
    }
    \caption{Block diagram of PQA, the proposed custom hardware implementation for PQ.}
    \label{fig:hw_diagram}
    \vspace{-5mm}
\end{figure*}

\subsection{PQA Overview}

Rather than relying on multiply-accumulate operations to calculate the result of a matrix-matrix multiplication, PQ performs memory lookups of precomputed results.
This requires designing hardware that can perform these lookups efficiently. 
Luckily, PQ has many hardware-friendly properties that can make a custom hardware design efficient. 
(1) Subspaces are completely independent, allowing us to partition the large dot product table (LUT$_{\text{PQ}}$) memory across multiple small on-chip memories to allow parallel memory accesses. 
(2) The compute-heavy distance computation---to find the closest prototype---needs to be done only once regardless of the number of output channels.
(3) With the exception of distance calculation, all other operations do not require heavy computations as they are either memory lookups or accumulations.

Figure~\ref{fig:hw_diagram} shows the architecture of PQA: our novel compute engine that can be used to perform PQ inference. The engine consists of 3 main modules:
A \textbf{distance calculation} module is responsible for determining the indices of the prototypes that are closest to the corresponding inputs.
Next, the \textbf{product lookup} module consists of a partitioned memory array that is responsible for looking up the product of the closest prototype and the current weight for each subspace.
Finally, the \textbf{accumulator} is responsible for adding results from different subspaces and producing the final outputs.
Note that we implement PQA on an Intel Agilex DE10 board with DDR4 external memory  with 36~GB/s transfer speeds, but we extrapolate our results to high-bandwidth memory (HBM) that can reach up to 460~GB/s using our hardware-verified cycle-accurate simulator that we develop \revision{based on performance model explained in Section}~\ref{sec:performance_modeling}. 
We provide more analysis of HBM in Section~\ref{sec:exp_evaluation}.

\subsection{PQA Hardware Components}

PQA is organized into processing lanes, replicated $N_\text{s}^{vec}$ times to process subspaces in parallel.
It is also able to produce $N_\text{out}^{vec}$ outputs at a time by further breaking down the large LUT$_{\text{PQ}}$ into smaller memories as shown in Figure~\ref{fig:hw_diagram}.
This allows producing multiple outputs at the same time.
As for inputs, the distance calculation module compares the input with $N_\text{p}^{vec}$ prototypes at a time where it compares $L_\text{s}^{vec}$ elements of the prototype each cycle.
The Compute Engine receives $L_s^{vec}$ portions of the input every cycle to be compared to all prototypes in each subspace, and to find the closest one depending on the distance function $d(\cdot)$. 
We implement both the L1 and L2 distance functions in PQA.
The index of the closest prototype is then sent to the product lookup portion that iterates over all weights, looks up the precomputed partial dot products, and accumulates these to compute each output.
In addition to the \textit{vectorization} parameters $L_\text{s}^{vec}$, $N_\text{p}^{vec}$, $N_\text{s}^{vec}$, $N_\text{out}^{vec}$, PQA contains \textit{maximum} parameters that are primarily used for \ommission{buffer sizing}\revision{sizing the input and LUT$_{PQ}$ buffers that hold layer values on-chip to enable single-cycle memory fetches during that layer's execution}: $L_\text{s}^{max}$, $N_\text{p}^{max}$, $N_\text{s}^{max}$, $N_\text{out}^{max}$, and $N_\text{in}^{max}$.

\subsubsection{Distance Calculators} 
This module consists of many processing lanes, one per $N_s^{vec}$. 
The lane contains $N_p^{vec}$ difference calculators. 
Each difference calculator corresponds to a certain prototype. The lane receives an input which is $L_s^{vec}$ elements wide, passes it to all difference calculators, each producing the difference between this input and its corresponding prototype. These differences go to a comparator which is responsible for outputting the index of the prototype with the minimum distance from the input vector. 
The comparator needs to cache the minimum distance so far to compare it with upcoming comparisons in the next clock cycle with the next set of prototypes since the hardware is vectorized over $N_{p}^{vec}$ to only perform some of the comparisons in each cycle.

Each difference calculator contains one group of prototypes $N_{p}$ and is responsible for calculating the distance between the input and these specific prototypes. 
It does that by passing each of the input elements along with its corresponding element in the prototype into the difference module then the output of all difference modules is added to an accumulator to produce the total distance after processing all $L_\text{s}^{vec}$ portions of the input. 

\subsubsection{Product Lookup and Accumulators} 
The product table memory is partitioned into $N_s^{vec}$ different banks to allow parallel lookups for each subspace. 
Furthermore, each subspace's LUT is partitioned into $N_{out}^{vec}$ different memories---each handles a different set of output channels---to further parallelize product lookup and to produce multiple outputs simultaneously.
PQA is therefore able to perform $N_{out}^{vec}\times N_s^{vec}$ LUT$_{\text{PQ}}$ lookups in parallel. This custom on-chip memory hierarchy is key to PQA's acceleration, and is not achievable with commodity hardware.
Finally, results are accumulated across subspaces using a set of parallel accumulators repeated $N_{out}^{vec}$ times. 
The accumulator is needed (and not a simpler adder tree) because of the vectorization along $N_s^{vec}$ which means that the difference calculated at each cycle might only be a partial dot product until all subspaces are processed.

\subsection{Numerical Bitwidths for PQA Components}

To further improve efficiency, lower numerical bitwidth can be used for PQ operations as explained in Section~\ref{sec:pq_quantization}.
We parameterize our hardware so that it can operate with any bitwidth in both the distance calculation and product lookup portions---each of those parts can use a different bitwidth since they are only connected with the index of the closest prototype, used to form the address for the LUT$_{\text{PQ}}$ memories. 
The output of LUT$_{\text{PQ}}$ is dequantized to 16 bits before accumulation, then subsequently, the output of the accumulators is quantized again before storing in the input buffer.
This ensures that that all buffers benefit from the smaller bitwidths, while using a wider bitwidth for accumulation.

\subsection{Performance Modeling}
\label{sec:performance_modeling}

Because of the vectorization along the $L_\text{s}^{vec}$, $N_\text{p}^{vec}$, $N_\text{s}^{vec}$, and $N_{out}^{vec}$ dimensions, a simple estimate of the total number of compute cycles per layer can be computed using Equation~\ref{eq:compute}.

\begin{equation}
    Cycles_{compute} = 
    \max\bigg(\left\lceil \frac{N_\text{p}}{N_\text{p}^{vec}} \right\rceil \times 
    \left\lceil \frac{L_\text{s}}{L_\text{s}^{vec}} \right\rceil, 
    \left\lceil \frac{C_\text{out}}{N_\text{out}^{vec}} \right\rceil\bigg) \times
    \left\lceil \frac{N_\text{s}}{N_\text{s}^{vec}} \right\rceil \times \frac{WH}{s^2}
    \label{eq:compute}
\end{equation}

There are 2 compute stages: Distance calculation, taking $\left\lceil \frac{N_\text{p}}{N_\text{p}^{vec}} \right\rceil \times 
\left\lceil \frac{L_\text{s}}{L_\text{s}^{vec}} \right\rceil$ cycles, and product lookup, taking $\left\lceil \frac{C_\text{out}}{N_\text{out}^{vec}} \right\rceil$ cycles, the maximum of those 2 determines the total compute taken to process the subspace. Since $N_s^{vec}$ subspaces are processed in parallel, we multiply that maximum by $\left\lceil \frac{N_\text{s}}{N_\text{s}^{vec}} \right\rceil$. Finally, this process is repeated for each of the $\frac{WH}{s^2}$ columns of the input.
In parallel to the compute cycles, memory loading of prototypes and $LUT_{PQ}$ occurs in parallel. The number of cycles needed for that can be computed using Equation~\ref{eq:load}. 

\begin{equation}
    Cycles_{load} = 
    \max \bigg(\left \lceil \frac{C_{out} \times N_p \times N_s}{N_{out}^{vec} \times N_s^{vec}} \right \rceil, \left \lceil \frac{|\mathbf{B}_{l}| + |\text{LUT}_{PQ}|}{\text{Mem}_{BW}} \right \rceil \bigg) 
    \label{eq:load}
\end{equation}

\noindent where $|\mathbf{B}_{l}|$ and $|\text{LUT}_{PQ}|$ are the size in bits of all prototypes and $\text{LUT}_{PQ}$ for each layer. 
$\text{Mem}_{BW}$ is the external memory bandwidth in bits/s. 
Equation~\ref{eq:load} takes a maximum between the memory loading cycles from the external memory bus, and the internal memory bandwidth of $\text{LUT}_{PQ}$. 
In most cases the internal memory bandwidth is much higher than external, but for some configurations of PQA with HBM memory \revision{where the value of $Mem_{BW}$ is much higher}, the situation may be flipped, \revision{where the internal memory bandwidth becomes the bottleneck,} thus necessitating the max operator in our analytical model.

\section{Experimental Setup}
\label{sec:exp_setup}

Here we provide a detailed description of the hardware environments, models, dataset, hyperparameters and training schemes used in our per-layer study and full-model results.
We also explain our enhancements to training PQ-DNNs to minimize accuracy degradation.

\subsection{HW-Setup}
\label{sec:exp_setup_HS}

We designed PQA using Intel's OpenCL SDK 21.2, targeting the DE10-Agilex FPGA board with the AGFB014R24B2E2V FPGA. 
Throughout our results, we compare PQA to a canonical systolic array-based deep learning accelerator (DLA) for native convolutions~\cite{aydonat2017opencl, dla_new}.
One of the important considerations when designing a custom hardware is the area it requires. 
Area and Fmax shown in the study are based on a real hardware accelerator running on the FPGA using the Intel OpenCL runtime. 
We express the area in terms of the number of equivalent Adaptive Logic Modules (eALMs) by following prior work that quantified the area of one DSP to be equivalent to 30 ALMs, and one BRAM is equivalent to 40 ALMs~\cite{rashid2014comparing}\revision{\footnote{\revision{While prior work uses a different family of FPGAs, it is the best available public estimate to the best of our knowledge and we believe it is sufficient for our evaluation.}}}.
For number of cycles, we verified that the numbers resulting from our analytical model described in Section~\ref{sec:performance_modeling} with the measured hardware performance measured by running on the DE-10 Agilex board. 
We used the analytical performance model in some of our results, especially to simulate HBM memory instead of the DDR4 that is available on the DE-10 board.
We compare PQA to a canonical systolic array-based deep learning accelerator (DLA) for native convolutions~\cite{aydonat2017opencl} running on an Arria 10 FPGA (20~nm technology node). 
For a fair comparison with our 8-nm Agilex PQA, we scale the reported DLA frequency by 1.6$\times$ to account for the newer process technology.
\revision{We get this scaling factor by comparing the frequency of a finite impulse response (FIR) filter design (1.60$\times$ frequency scale factor), and a Fast Fourier Transform (FFT) design (1.47$\times$ frequency scale factor) when implemented on the Arria 10 versus the Agilex FPGA, and we opted to use the higher scaling factor to favor the baseline DLA.}

\comment{
We get this scaling factor by comparing the frequency \revision{of} a finite impulse response (FIR) filter design when implemented on the Arria 10 versus the Agilex FPGA. \revision{To further validate this scale factor, we found the frequency of a Fast Fourier Transform (FFT) design on both devices and the ratio was 1.47$\times$ which is lower than our chosen scaling factor.}
}

\subsection{Models \& Training}
\label{sec:models_training}

\subsubsection{Datasets}
\label{app:datasets}
We make use of two image classification datasets: CIFAR-10~\cite{krizhevsky2009learning} and EMNIST~\cite{emnist}. The former is comprised of 50K $32\!\times\!32$ RGB images for training and 10K for testing, with both sets evenly split along ten image classes. The EMNIST dataset on the other hand is much larger totaling 112,800 and 18,800 images for training and testing respectively. We use the \textit{balanced} partitioning of EMNIST which contains $28\!\times\!28$ greyscale images of digits and letters resulting in 47 classes with, as the name suggests, the same number of examples. The current best performing architecture on this dataset reaches 91.06\%~\cite{wavemix} in this partition. For both training sets we randomly leave 10\% out for validation. For keyword spotting we rely on the SpeechCommands~\cite{speechcommands} data which is comprised of 105,829, 16-KHz 1-second long audio clips of a spoken word (e.g. "yes", "up", "stop") and the task is to classify these correctly into 12 possible classes. Similar to previous works~\cite{zhang2018hello,berg21_interspeech,MLSYS2021_a3c65c29} we pre-process each audio clip and extract 10 MFCC~\citep{mfcc} features using a 40ms window with a 20ms stride resulting in 10$\times$49 input matrices.

\subsubsection{Training Infrastructure}

\begin{table*}[]
    \centering
        \caption{Hyperparameters used in our experiments. Due to the large space of PQ configurations and $\tau$ values considered, there is not a single \textit{golden config} for each model. Here we report the ranges of hyperparemters that we found to deliver good quality models across PQ settings. Column \textit{Early Stop $\tau$} indicate the epoch from which temperature $\tau$ is kept fixed for the remaining of the training.}
    \scalebox{0.8}{
        \begin{tabular}{lcccccccc}
        \toprule
        \textbf{Model} & \textbf{Epochs} &  \textbf{Batch} & \textbf{LR (params)} & \textbf{LR (proto)} & \textbf{Start $\tau$} &  \textbf{End $\tau$} &  \textbf{Early Stop $\tau$} & \textbf{Scheduler}\\
        \toprule
        ResNet20 & 120 & 64 & [0.005,...,0.05] & [0.025,...,0.075] & 1.0 & 0.0005& 10  & StepLR [40,60,90] \\
        \midrule
        MicroNet-PQ & 30 & 96 & [0.001,...,0.01] & [0.001,..., 0.025] & 1.0 & [0.0005, 0.001] & 10 & ExponentialLR [0.25,...,2.0] \\
        \midrule
        DW$_{\text{EMNIST}}$ & 90 & 96 & [0.000,...,0.001] & [0.005,...,0.05] & 1.0 & 0.0005 & 10 & StepLR [30,50,70]\\
        \bottomrule
        \end{tabular}
    }
    \label{tab:hparams}
\end{table*}
We implement our training infrastructure with PyTorch~\cite{pytorch2019}. We analyze 3 models: ResNet-20\cite{He_2016_CVPR} on CIFAR10, MicroNet\cite{MLSYS2021_a3c65c29} on KWS and a custom CNN with 10 depth-wise separable layers called DW on EMNIST. Full details of the model architectures is in Appendix~\ref{sec:model_arch}.
Training PQ models is difficult and is currently only proven to work on smaller models like the ones we consider. This is similar to other work in the literature\cite{pecan}.
We use two optimizers: one for the bank of prototypes in each layer $\mathbf{B}_{l}$; and another for the rest of the parameters in the model (e.g. the layer weights, and non-PQ layers.). Both instantiate an Adam optimizer~\cite{adam_optim} albeit with different learning rate. There are also two learning rate schedulers, each with its own decaying coefficient and scheduling. The hyperparameters for each model are presented in Table~\ref{tab:hparams}. We found learning rates and scheduling parameters to have large impact, not only in final model quality, but also in terms of training stability.

\subsubsection{PQ Training Enhancements}

A number of training techniques and tricks were considered and introduced to our training pipeline. We found gradient clipping to be crucial when training deeper models (\textit{i.e.} ResNet20 and DW$_{\text{EMNIST}}$). Value-based gradient clipping~\cite{Zhang2020Why} with small thresholds (\textit{e.g.} 0.25, 0.5) alleviated exploding gradients in some cases. During the first epochs of training, the encoded input is obtained with $\tau$ values that are gradually decreased. As it becomes smaller, the construction of the \textit{prototyped} input gets close to one-hot, impacting the flow of gradients. We implemented two mechanisms to counteract this while still ensuring the performance of one-hot encoding is not affected: formulating the prototype selection via a Gumbel-Softmax~\cite{gumbel}. This, however didn't seem to have a clear impact on the quality of our training. What did have a small positive impact was to introduce a stochastic masking to the final encoded input $\mathbf{\hat{X}^{\text{enc}}}_{n,j}$ prior to multiplying it with the layer weights. This masking, applied at the subspace level and individually to input column, allowed a fraction $\rho$ of vectors to remain unencoded. We found that small $\rho\!=\!0.1$ lead to higher final one-hot PQ accuracy. Higher values would prevent layer weights and prototypes to adequately operate in one-hot scenarios, as it is the case during inference.
Furthermore, we add a new term to our Cross Entropy training loss that encourages prototypes within a subspace to be orthogonal to each other~\cite{orthogonal}. 
We only found this regularization term to be useful to lessen the impact of sub-optimal training hyperparameters.

\section{Product-Quantizability: Study \& Trade-offs}
\label{sec:pq_study}

This section investigates PQ parameterizations and training enhancements, and the corresponding tradeoffs in PQ-DNN accuracy, compute, and memory footprint.

\subsection{PQ Implementation Trade-Offs}
\label{sec:PQ_tradeoffs}

Different parameterizations of PQ (\textit{i.e.} choice of \{$N_{\text{p}}$,\revision{$ L_{\text{s}}$}\}), will lead to dramatically different levels of model acceleration, memory footprint, and accuracy degradation. Understanding these trade-offs is fundamental to the design of PQ-DNNs.

\subsubsection{Compute footprint} 
The number of FLOPs to perform an {\tt im2col}-equivalent convolution (as in Figure~\ref{fig:tinypq_diagram}) is given by $\text{FLOPs}_{\tt{im2col}}\! =\! 2K^2C_{\text{in}}WH C_{\text{out}}$, assuming groups and stride equal to one and squared kernels. The same layer but implemented with PQ would result in $\text{FLOPs}_{\text{PQ}}\! =\! \text{FLOPs}^{\text{enc}}\! +\! \text{FLOPs}^{\text{add}}\!$. The first term is given by $\text{FLOPs}^{\text{enc}}\! =\!  N_{\text{s}}d(\cdot)_{\text{FLOPs}}WH$, with $d(\cdot)_{\text{FLOPs}}$ representing the FLOPs to compute the distances $\mathbf{d}^{(n)}_j$. Assuming Euclidean distance, $d(\cdot)_{\text{FLOPs}}\! =\! 3N_{\text{p}}L_{\text{s}}$. Term $\text{FLOPs}^{\text{add}}\! =\! (N_{\text{s}}-1)WHC_{\text{out}}$ accounts for the cost of performing the addition of partial dot-products needed to complete a full row-column dot-product $\mathbf{\hat{X}}\!\cdot\!\mathbf{\hat{W}}$. The FLOPs savings ratio is given by

\vspace{-1mm}

\begin{equation}
  \frac{\text{FLOPs}_{\tt{im2col}}}{\text{FLOPs}_{\text{PQ}}} \!=\! \frac{2K^2C_{\text{in}}WHC_{\text{out}}}{N_{\text{s}}WH(d(\cdot)_{\text{FLOPs}}\!+\!C_{\text{out}})} \!=\!  \frac{2C_{\text{out}}L_{\text{s}}}{3N_{\text{p}}L_{\text{s}} \!+\! C_{\text{out}}}
\label{eq:flops_footprint}
\end{equation}
\vspace{-1mm}

with $N_{\text{s}}\! =\! K^2C_{\text{in}}/L_{\text{s}}$ and simplifying $(N_{\text{s}}\!-\!1)$ as just $N_{\text{s}}$. This expression suggests that reducing the number of prototypes has a larger impact than increasing their length.\ommission{ See Figure~}\ommission{for a quantitative evaluation.} \revision{This can also be seen in Figure~}\ref{fig:overall_trend_acc_vs_size_of_lut} \revision{where the grey-shaded squares indicating the FLOPs increase heavily as you increase $N_p$ but they do not increase as much when you use larger $L_s$}.

\begin{table}
    \centering
    \caption{The best performing models in PECAN-D require a high number of prototypes ($N_{\text{p}}$) of short length ($L_{\text{s}}$). As a result, look-up tables are many times larger than the original non-PQ model.}
    \scalebox{1}{
    \begin{tabular}{lccccc}
        \toprule
        \textbf{Model} & \textbf{Params} & \textbf{$\text{N}_{\text{p}}$} & \textbf{$\text{L}_{\text{s}}$} & \textbf{$|\text{LUT}_{\text{PQ}}|$} & \textbf{Mem. Footprint} \\
        \midrule
        LetNet & 61K & 64 & 9 \& 8 & 489K & 8.0$\times$ \\
        ResNet20 & 269K & 128 \& 64 & 3 \& 4 & 5.7M & 21.3$\times$ \\
        \bottomrule
    \end{tabular}
    }
    \label{tab:pecan_footprint}
\end{table}

\subsubsection{Memory footprint} 
The number of parameters in a convolutional layer is given by $ C_{\text{out}}C_{\text{in}}K^2$, assuming squared kernels and groups=1. When transformed into PQ, this number becomes $\text{params}_{\text{PQ}}\! =\! |\mathbf{B}_{l}|\! +\! |\text{LUT}_{\text{PQ}}|\! =\! N_{\text{s}}N_{\text{p}}(L_{\text{s}}\! +\! C_{\text{out}})$. In this way, and assuming that $C_{\text{out}}\! \gg\! L_{\text{s}}$, savings in parameter count is possible when $C_{\text{in}}K^2/N_{\text{s}}N_{\text{p}}\!=\!L_{\text{s}}/{N_{\text{p}}}\!>\! 1$,
i.e., as it was the case when assessing the compute footprint of PQ, longer prototypes and few of them are also preferred.

\subsubsection{Accuracy degradation} 
Previous attempts of applying PQ to the entire network required using a larger number of short prototypes to maximise model accuracy. In PECAN-D~\cite{pecan}, this translated into a very large increase in memory footprint, see Table~\ref{tab:pecan_footprint}. While intuitively lower $L_{\text{s}}$ should always be preferred when prioritising accuracy degradation, it is unclear what the underlying trade-offs between $N_{\text{p}}$ and $L_{\text{s}}$ are at different layers of a network.

\begin{figure}[t]
    \centering
       \includegraphics[width=0.7\linewidth]{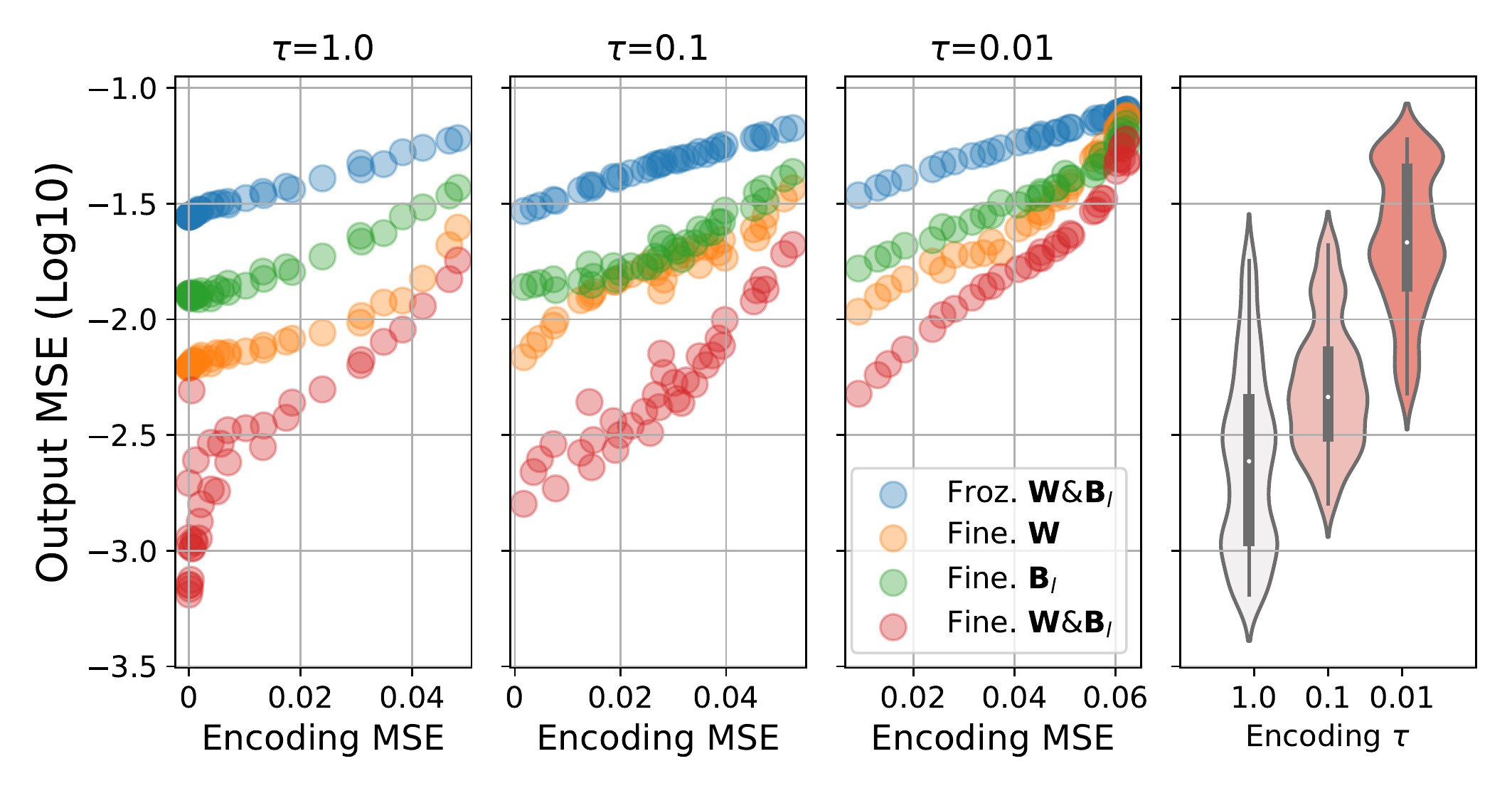}
    \vspace{-0.5cm}
    \caption{As the temperature $\tau$ decreases, the error in the output of a PQ layer increases rapidly. Even when prototypes and layer parameters are finetuned to minimise MSE($\mathbf{Y}_{\text{PQ}}$,$\mathbf{Y}$), most of the \{$N_{\text{p}}$,$ L_{\text{s}}$\} configurations (each configuration is a dot) leads to much larger error (see rightmost subplot).}
    \label{fig:overall_trend_acc_vs_encoding_mse}
\end{figure}

\begin{figure}
    \centering
       \includegraphics[width=0.6\linewidth]{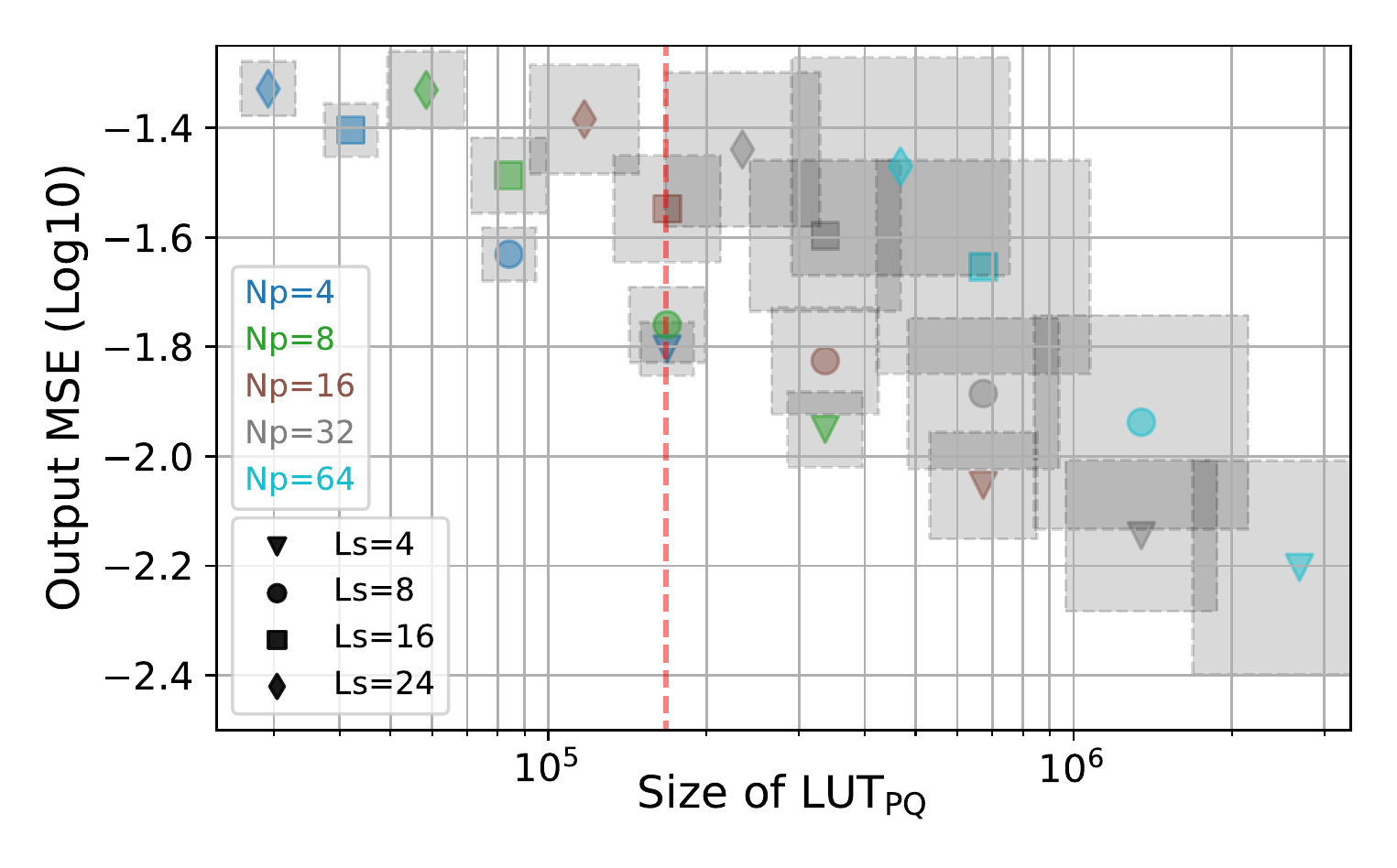}
       \vspace{-0.5cm}
    \caption{Analysis of a layer with a $366\!\times\!4\!\times\!4$ input. When spatial dimensions of the input are small relative to their channel dimension, fewer prototypes lead to similar error to other configurations that result in larger $\text{LUT}_{\text{PQ}}$ and a higher $\text{FLOPs}^{\text{enc}}$ (area of grey-shaded squares) as in Eq.~\ref{eq:flops_footprint}. The red line indicates the size in the equivalent non-PQ layer.}
    \label{fig:overall_trend_acc_vs_size_of_lut}
\end{figure}

\subsection{PQ Layer-wise Parameter Sweeps}
\label{sec:pq_sweeps}

In this section we assess the impact of different \{$N_{\text{p}}$,$ L_{\text{s}}$\} in terms of memory and compute footprint as well as accuracy degradation. 
We design a per-layer analysis where the task for each PQ layer is to generate an output $\mathbf{Y}_{\text{PQ}}$ that is as close as possible to $\mathbf{Y}$, the output of an equivalent layer from a standard, non-PQ, pretrained model. 
Framing this empirical study in such way enables us to isolate individual layers from the impact of other elements in the model, training hyperparameters, and dynamics. 
This study uses DW$_{EMNIST}$ a CNN containing 10 depth-wise separable convolutional layers designed for image classification on the EMNIST~\cite{emnist} dataset as discussed in Section~\ref{sec:models_training}\revision{.} %

Given a PQ layer with randomly initialised $\mathbf{B}_{l}$ and weights from its non-PQ counterpart, %
a two-phases study is conducted. First, the prototypes are trained to minimise $\text{MSE}^{\text{enc}}(\mathbf{\hat{X}^{\text{enc}}}, \mathbf{X})$; then, the divergence of $\mathbf{Y}_{\text{PQ}}$ w.r.t. $\mathbf{Y}$ is measured in four different scenarios, namely when every other element in the layer is kept frozen (blue dots\revision{)} in top plot of Figure~\ref{fig:overall_trend_acc_vs_encoding_mse}, when prototypes are further finetuned to minimise $\text{MSE}^{\text{out}}\!=\!\text{MSE}(\mathbf{Y}_{\text{PQ}},\mathbf{Y})$ (green dots), when only the layer weights are finetuned (orange dots), and when both prototypes and layer weights are jointly finetuned (red dots). All settings involving finetuning start from the same state of trained prototypes (blue dots). These two phases are repeated for a broad \{$N_{\text{p}}$, $ L_{\text{s}}$, $\tau$\} range and across all layers. We make the following observations.

\noindent\textbf{1. Learning one-hot encodings is hard.} As temperature parameter $\tau$ decreases, there is a shift in the distribution over \{$N_{\text{p}}$, $ L_{\text{s}}$\} settings that lead similar $\text{MSE}^{\text{enc}}$ and $\text{MSE}^{\text{out}}$ trends (blue, green orange dots in Figure~\ref{fig:overall_trend_acc_vs_encoding_mse}). More evidently (red dots and violin plot) is the impact on $\text{MSE}^{\text{out}}$ when both layer weights and prototypes are finetuned. These results suggest that at least certain \{$N_{\text{p}}$, $ L_{\text{s}}$\} can perform well at lower $\tau$ values but most cannot.%

\noindent\textbf{2. Reasons for larger $\mathbf{\textit{N}}_{\text{p}}$ decrease with depth.} %
Accuracy decreases faster by lowering $L_{\text{s}}$ than increasing $N_{\text{p}}$ as can be observed in Figure~\ref{fig:overall_trend_acc_vs_size_of_lut}. However, this dimension brings a large increase in memory footprint since the size of $\text{LUT}_{\text{PQ}}$ is inversely proportional to $L_{\text{s}}$ (\textit{i.e.} shorter prototypes lead to more subspaces, which in turn leads to more pre-computed dot products). In comparison, larger $N_{\text{p}}$ has a lesser impact on $\text{MSE}^{\text{out}}$ overall but affects both $\text{LUT}_{\text{PQ}}$ size and $\text{FLOPs}^{\text{enc}}$ linearly. Because of this reason, trading  $N_{\text{p}}$ for $L_{\text{s}}$ is desirable even at the cost of some accuracy degradation (\textit{e.g.} \{$N_{\text{p}}\!=\!4$,$L_{\text{s}}\!=\!4$\} vs \{$N_{\text{p}}\!=\!32$,$L_{\text{s}}\!=\!8$\}, the latter resulting in just a $1.13\times$ lower error.). %

\begin{figure}[t]
    \centering
  \includegraphics[width=.6\linewidth]{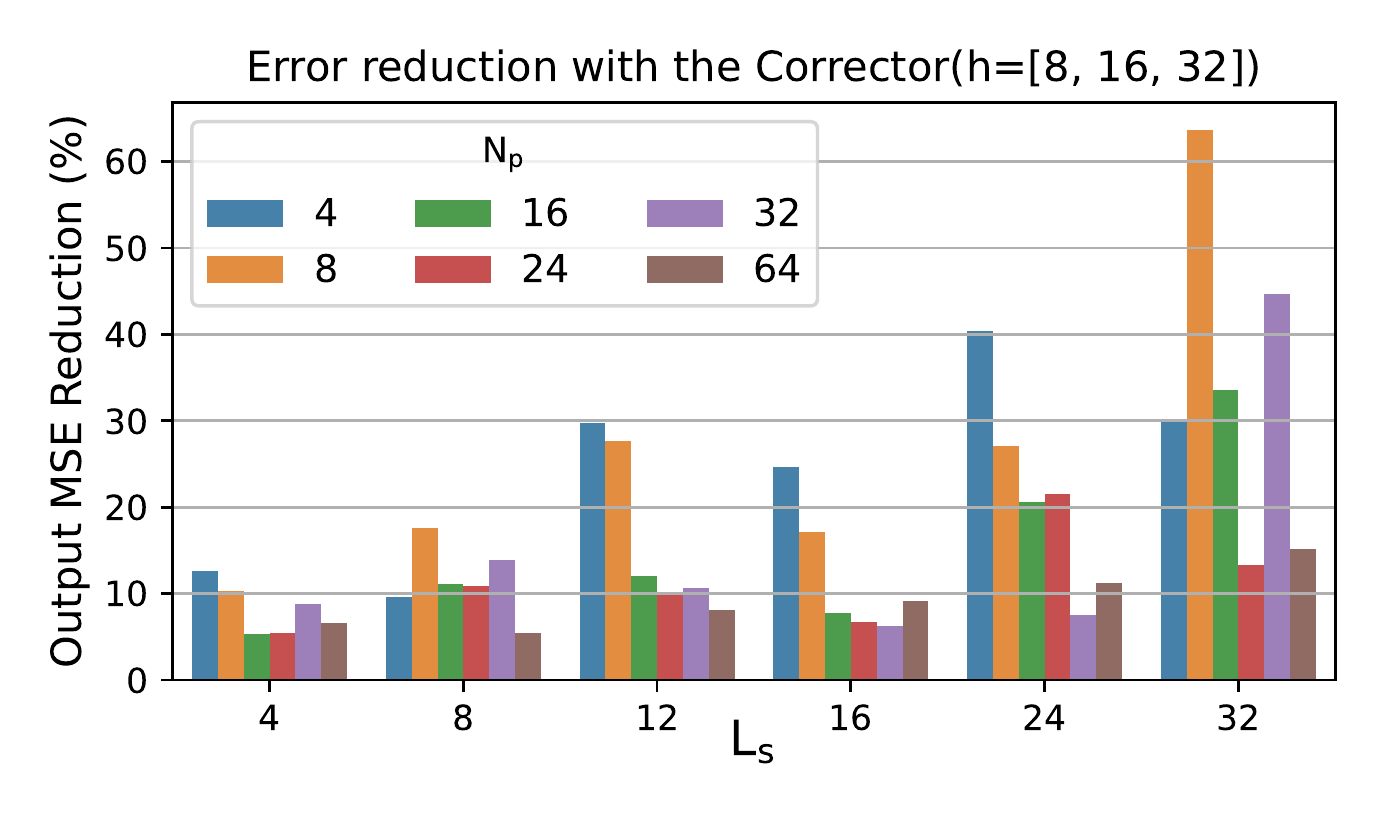}
  \vspace{-7mm}
    \caption{Maximum observed error reduction with corrector: a 2-layer MLP with hidden dimension $h$. %
    Longer and fewer prototypes benefit more on average.}
    \label{fig:corrector_error}
\end{figure}

\noindent\textbf{3. Amortizing and reducing encoding costs.} Computing the distances $\mathbf{d}^{(n)}$ between input columns and prototypes in the $n$-th subspace might come at a too high cost given that PQ only makes use of the index of the \textit{closest} prototype to each column and not its actual value. A less computationally demanding distance could be used. 
Alternatively, we could further leverage $\mathbf{d}^{(n)}$ and better amortise $\text{FLOPs}^{\text{enc}}$. To this end, we design a lightweight MLP \textit{corrector} that, given these distances, it learns a transformation that helps further reducing $\text{MSE}^{\text{out}}$. Intuitively, the corrector has a higher potential for correction with longer and fewer prototypes. This is the pattern in Figure~\ref{fig:corrector_error} where we show the maximum observed reduction in $\text{MSE}^{\text{out}}$ across all layers in the study. 
Even though a well tuned corrector can still benefit PQ layers with short prototypes, its computational footprint for a given hidden dimension $h$, dominated by $N_{\text{s}}N_{\text{p}}\!\times\!h$, does not justify its use.
Therefore, we do not use the concept of a correct MLP in this work and we defer to future work to build on this idea.

\begin{figure}[t]
\centerline{
\includegraphics[width=0.8\linewidth, trim = 0cm 0cm 0cm 0cm]{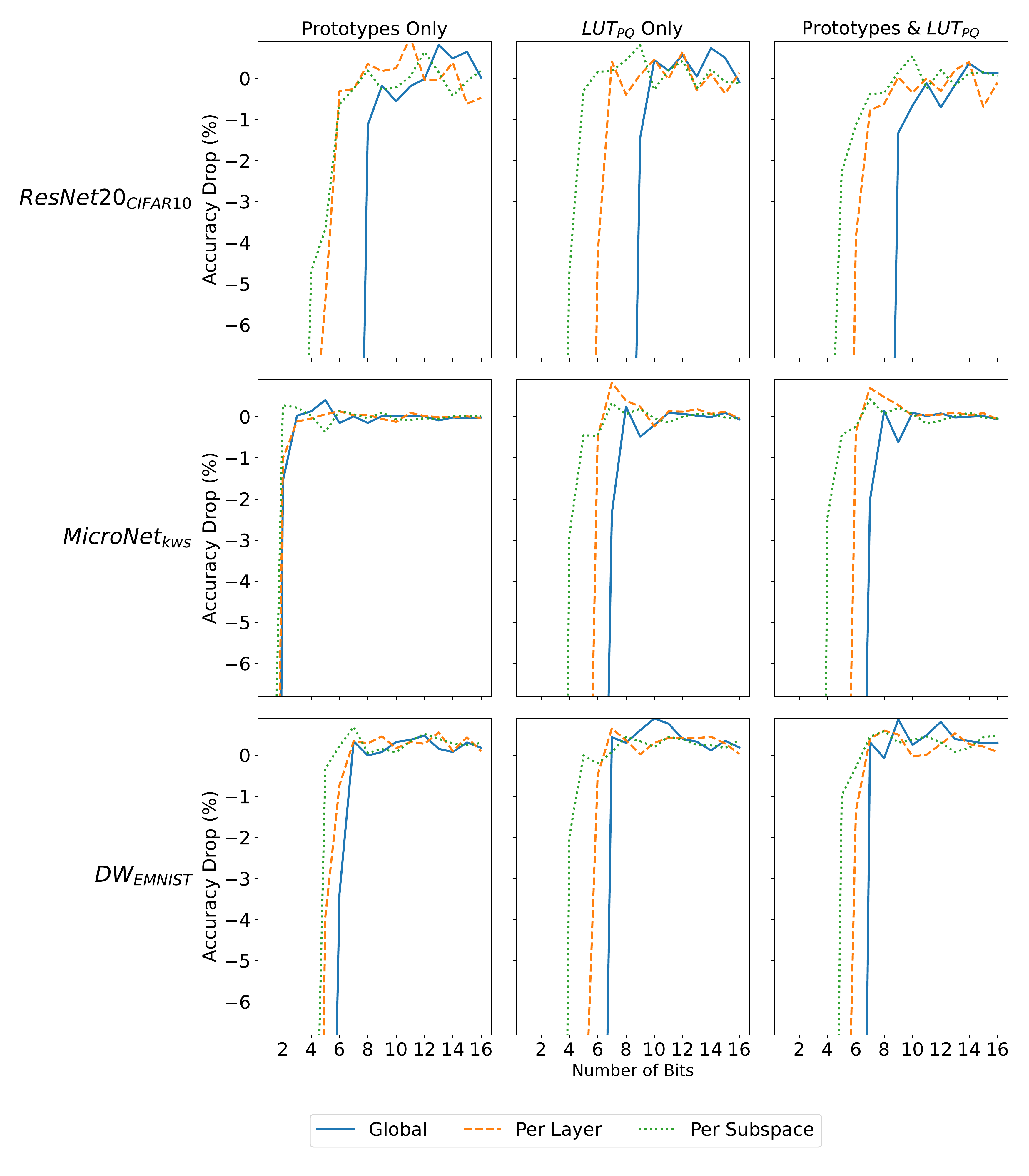}
}
\vspace{-0.3cm}
\caption{Drop in accuracy due to quantization using different techniques for multiple models. \textbf{Global} means that a single scale and offset are used for whole model. \textbf{Per Layer} and \textbf{Per Subspace} mean that the scale and offset are different for each layer and each subspace respectively.}
\label{fig:accuracy_plot}
\end{figure}

\subsection{Numerical Bitwidth Analysis}
\label{sec:quantization}

After training PQ-DNNs, we perform post-training quantization as described in Section~\ref{sec:pq_quantization}.
Figure~\ref{fig:accuracy_plot} illustrates the impact of decreasing the bitwidth from 16~bits down to 2~bits for the distance calculator, product lookup, and when reducing the bitwidth for both.
A validation set was used to identify the minimum and maximum values to dynamically compute the scale and offset values for quantization.
We achieved better accuracy when using the $30th$ and $70th$ percentile values instead of the full range for computing the scale and offset values for our integer quantization---this enhanced version has been applied to MicroNet$_\text{KWS}$ in Figure~\ref{fig:accuracy_plot}.
As Figure~\ref{fig:accuracy_plot} shows, PQ operations can tolerate very low bitwidths (2--5 bits, depending on the DNN), especially when per-subspace quantization is used. 
Prototypes are easier to quantize than LUT$_{\text{PQ}}$.
In the following evaluations, we tailor these bitwidths to each PQ-DNN, in addition to finding the ideal hardware vectorization parameters to maximize efficiency.

\section{Experimental Evaluation}
\label{sec:exp_evaluation}

This section investigates the performance and efficiency of PQA through both layerwise and end-to-end DNN execution.
To guage the efficiency of PQA, we compare to an optimized deep learning accelerator (DLA) from prior work~\cite{aydonat2017opencl,dla_new}.
In addition, we compare to the latest PQ work, PECAN~\cite{pecan}.
Our results demonstrate both performance and efficiency improvements compared to both PECAN and DLA, paving the way for further work to establish the utility of PQ for DNN efficiency.

\subsection{PQA Layerwise Performance Analysis}

\begin{figure*}[t]
    \centering
    \begin{subfigure}[b]{0.49\textwidth}
        \centering
        \includegraphics[width=\linewidth, trim = 0 0 1.5cm 0]{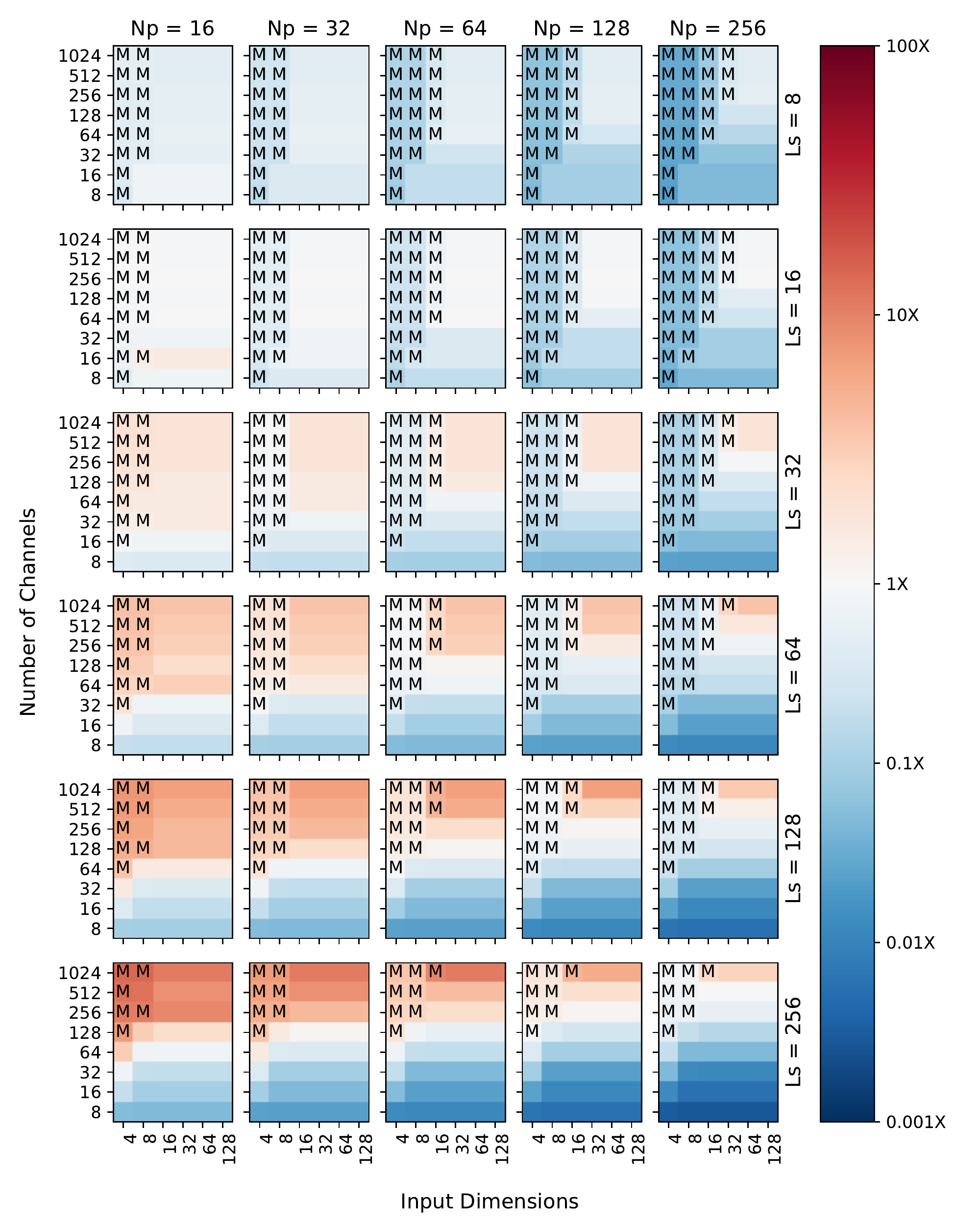}
        \caption{DDR4}
        \label{fig:speedup_on_different_inputs_and_params}
    \end{subfigure}
    \hfill %
    \begin{subfigure}[b]{0.49\textwidth}
        \centering
        \includegraphics[width=\linewidth, trim = 0 0 1.5cm 0]{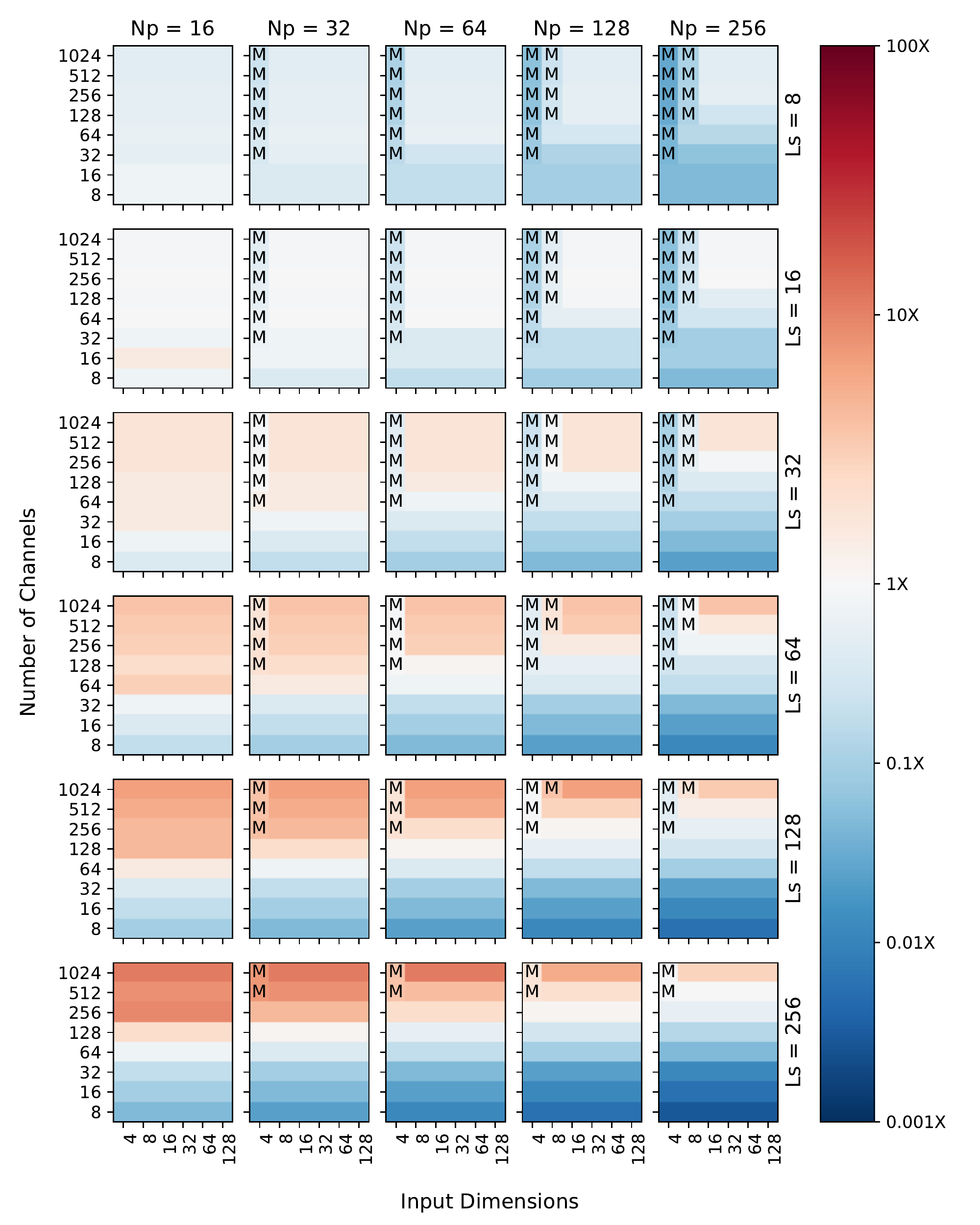}
        \caption{HBM}
        \label{fig:heatmap_3_hbm}
    \end{subfigure}
    \caption{Speedup of different $\{N_\text{p}, L_\text{s}\}$, input size and channels given a 3$\times$3 kernel on PQA relative to DLA at 16 bits, when using (a) DDR4 and (b) HBM. M = Memory Bound.}
    \label{fig:combined_heatmap}
\end{figure*}

Figure~\ref{fig:combined_heatmap} shows the speedup of running a convolution layer using PQ on a custom PQA compared to running its non-PQ equivalent on a custom DLA. %
In this study, all PQA vectorization parameters are set to $16$.
The convolution has a kernel of size 3$\times$3 and the number of input channels is assumed to be the same as the number of output channels. 

As expected, smaller $N_\text{p}$ leads to faster PQ execution as the cycles needed to compare prototypes are fewer. 
Larger $N_\text{p}$ increases the LUT$_{\text{PQ}}$ size and the time needed to load it from external memory, causing more points to become memory-bound. 
When using a High Bandwidth Memory (HBM) in Figure~\ref{fig:heatmap_3_hbm}, \ommission{some memory bound configurations become compute bound and in multiple configurations, a higher speedup is achieved.} \revision{many memory bound configurations are no longer memory bound, for example $N_p=16$ has no memory bound cells in HBM while more than half of the cells with input dimensions 4 and 8 are memory bound in DDR4. We can also see some of the cells that didn't result in a speedup in DDR4, resulting in a speedup in HBM because they are no longer memory bound. For example, in $L_s = 32$, the speedup area in all values of $N_p \geq 32$ in HBM includes smaller dimensions that didn't show a speedup in case of DDR4.}
When increasing $L_\text{s}$, the number of subspaces ($N_\text{s}$) decreases and speeds up overall execution---we observe speedups starting from $L_\text{s}\!=\!32$. \revision{We can see a single row in $L_\text{s}$=16 showing speedup which is the row that has the number of channels and $N_\text{p}$ set to 16 as well. This is because our vectorization parameters are all set to 16 making these dimensions utilize the hardware very efficiently unlike the surrounding cells that do not match the vectorization parameters perfectly.}
Finally, we can see that larger layers benefit more from PQ with speedups up to 100$\times$ at small $N_\text{p}$. 
In general, more aggressive PQ quantization leads to less computation and memory and therefore a better speedup. 
It is clear that running PQ on custom hardware is not always advantageous, especially when compared to an optimized DLA. 
However, we have shown that unlike CPUs and GPUs, we are able to find layer sizes and PQ parameterizations for which PQ can indeed outperform conventional convolutions. 
We have also shown that HBM alleviates memory bottlenecks and often favors PQA over DLA because of the reliance of PQ on relatively larger memory bandwidth.

\comment{
\begin{figure*}[t]
    \centering
    \begin{subfigure}[b]{0.48\textwidth}
        \centering
        \includegraphics[width=\linewidth]{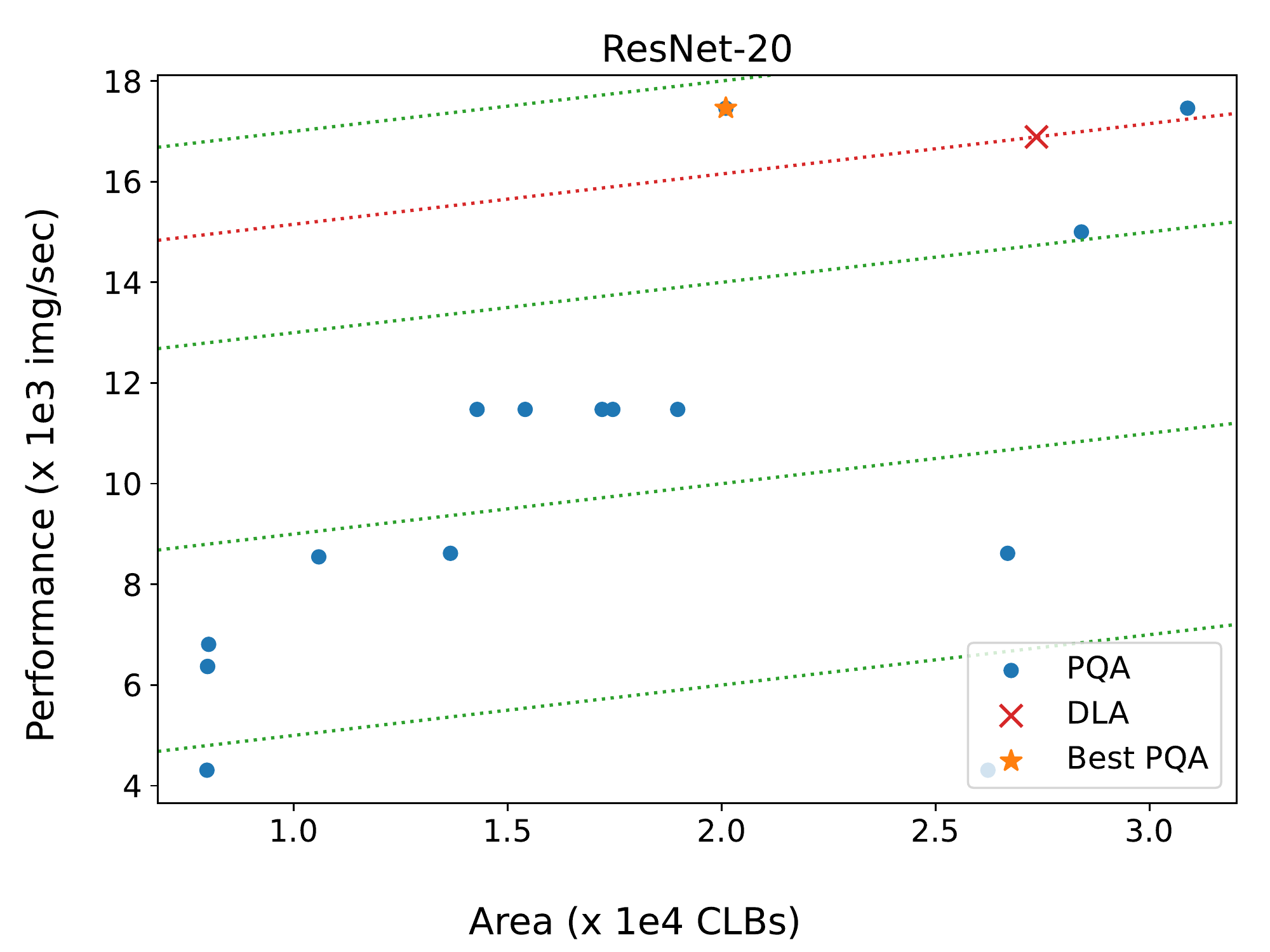}
        \caption{DDR4}
        \label{fig:hw_latency_vs_area}
    \end{subfigure}
    \hfill %
    \begin{subfigure}[b]{0.48\textwidth}
        \centering
        \includegraphics[width=\linewidth]{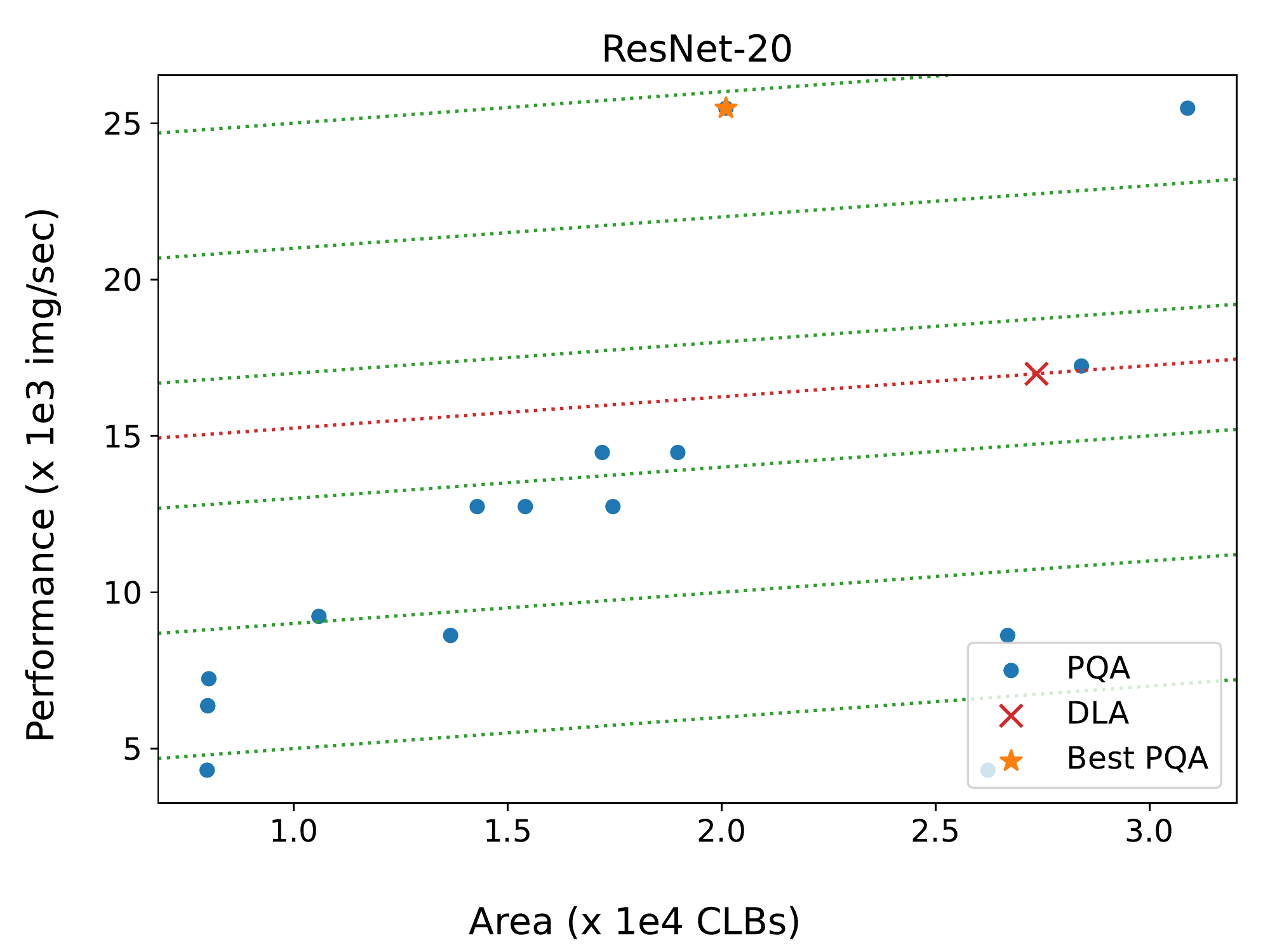}
        \caption{HBM}
        \label{fig:performance_vs_area_hbm}
    \end{subfigure}
    \vspace{-3mm}
    \caption{Performance vs area for different PQA for ResNet20. Best DLA configurations are plotted for reference. Dotted lines represent constant performance per area. Left uses DDR memory while right uses HBM memory. All PQA variants use 16-bit precision. \moh{x-axis should say 1e3 eALMs and remove "ResNet20" plot titles. also increase marker sizes}}
    \label{fig:combined_performance_area}
\end{figure*}

\moh{we may end up deleting the following paragraph and figure 11 altogether. The numbers are too imprecise and there isn't much we're getting from it that we don't get in the results table later.}
To further explore the relative performance and efficiency of PQA and DLA, we plot the end-to-end performance of ResNet20, and the corresponding accelerator areas in Figure~\ref{fig:combined_performance_area}.
This analysis exposes PQA configurations that outperform DLA for both DDR4 (Figure~\ref{fig:hw_latency_vs_area}) and HBM (Figure~\ref{fig:performance_vs_area_hbm}).
For Resnet20, the PQA parameters that achieves best performance at 16 bits are $N_{out}^{vec}=32$, $N_\text{s}^{vec}=16$, $N_\text{p}^{vec}=16$, $L_\text{s}^{vec}=16$.
However, it is important to note that many other configurations fail to match or outperform DLA, indicating the importance of codesigning PQA vectorization parameters with each DNN.
It is clear that using HBM further boosts the performance of PQA relative to DLA because of the increased memory bandwidth.
}

\subsection{MicroNet Case Study}
\label{sec:full_model_PQ_results}

\begin{figure*}[t]
    \centering
    {
    \includegraphics[width=\linewidth, trim=0 .8cm 0 0]{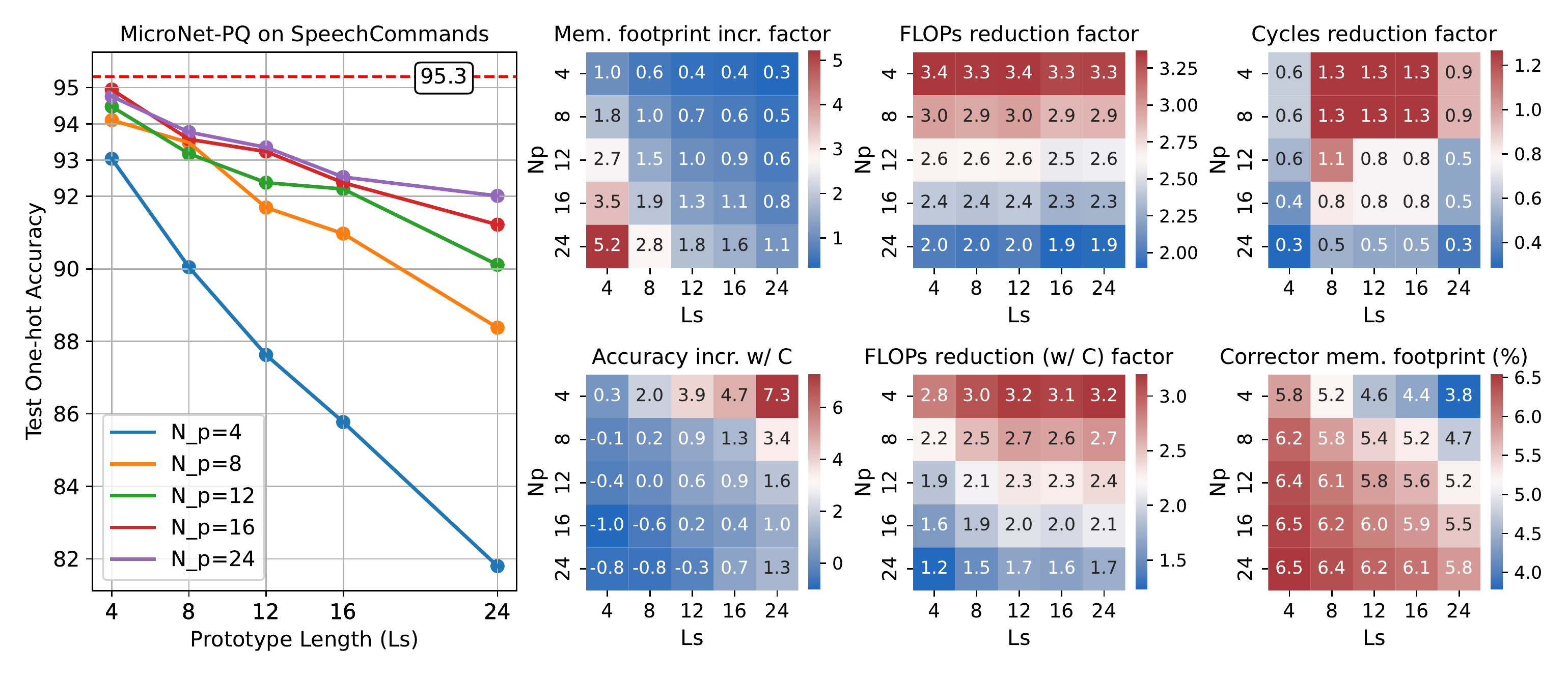}
    }
    \vspace{-4mm}
    \caption{MicroNet-PQ reaches 95.0\% accuracy with \{$N_{\text{p}}\!=\!16$,$L_{\text{s}}\!=\!4$\} leading to a $2.4\times$ reduction in the number of inference FLOPs. However, fewer FLOPs does not always translate \ommission{in}\revision{to} faster inference (cycles). When increasing memory footprint is not possible (\{$N_{\text{p}}\!=\!8$, $L_{\text{s}}\!=\!8$\}), Micronet-PQ still reaches over 93.5\% while resulting in nearly $3\times$ lower computational footprint. With the corrector (displayed as `w/C`), models with longer prototypes can reach that threshold accuracy too, at a small increase in memory footprint and FLOPs.}
    \label{fig:kws_results}
\end{figure*}

Figure~\ref{fig:kws_results} sheds light on the accuracy, compute, memory, and latency implications of PQ with different parameterizations of $L_s$ and $N_p$.
The line plot in Figure~\ref{fig:kws_results} (left) shows the performance of MicroNet-PQ under varying configurations with just a 0.3\% gap compared to the non-PQ baseline when \{$N_{\text{p}}\!=\!16$, $L_{\text{s}}\!=\!4$\}. 
The accompanying heatmaps highlight different trade-offs between MicroNet-PQ and their non-PQ counterpart. When the accuracy requirement is lower, longer prototypes (higher $L_{\text{s}}$) can be selected to achieve a better FLOPs speedup ratio. 
With the corrector introduced in Section~\ref{sec:pq_sweeps}, some of the accuracy drop introduced by longer prototypes can be recovered by up to 7.3\% at a small memory footprint increase.
Figure~\ref{fig:kws_results} highlights the theme of our analysis in Section~\ref{sec:hw_model}: FLOPs and memory footprint are not a good proxy for hardware latency for PQ as shown in the top 3 heatmaps of the figure.
Looking at the top row of heatmaps, we can see that a decrease in both FLOPs and memory footprint does not always result in a proportional decrease in hardware execution cycles.
However, many PQ configurations, with $N_{out}^{vec}$ set to 64, eventually lead to a speedup for \revision{PQA}, as shown in the \revision{top right heatmap in the figure where the cycle reduction factor reaches 1.3}.

\subsection{Area, Latency and Frequency Trends of PQA}

\begin{figure}[htbp]
\centerline{
\includegraphics[width=0.9\columnwidth, trim=0 .7cm 0 0]{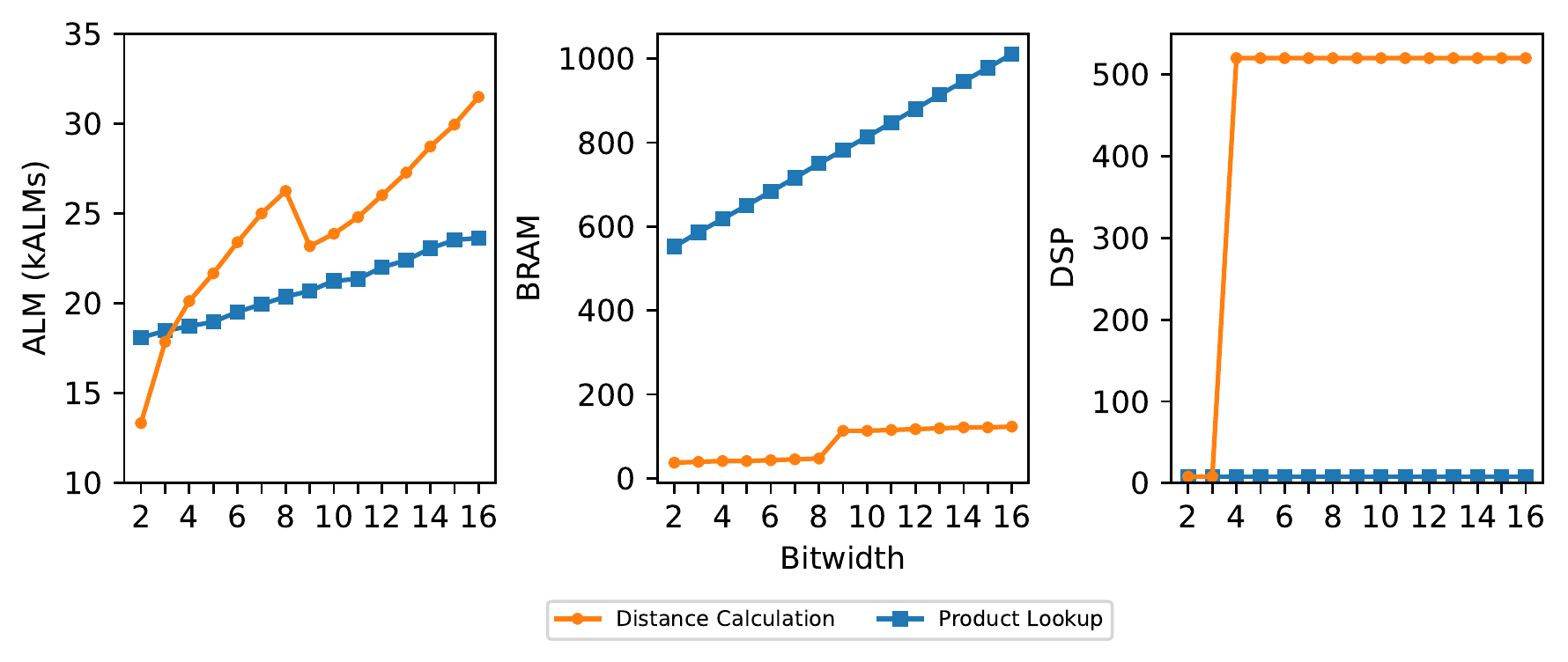}
}
\caption{Area trends of the distance calculation operation with varying prototype bitwidths and product lookup operation with varying $LUT_{PQ}$ bitwidths. All PQA vectorization parameters are set to 16.}
\label{fig:pq_area}
\end{figure}

We study the effects of varying the inputs/prototype and $LUT_{PQ}$ bitwidths on the two main PQ operations, distance calculation and product lookup, by sweeping bitwidths on a version of PQA for MicroNet. 
We used the PQ parameters $\{L_{s}$= 4, $N_{P}$=16$\}$ that optimized accuracy and performance on MicroNet, and we used the same values for the hardware vectorization parameters $\{L_{s}^{vec}$= 4, $N_{p}^{vec}$=16$\}$.
We choose the remaining vectorized parameters for performance considerations and on-chip buffer sizes according to the MicroNet$_{KWS}$ layer parameters \cite{banbyr2021micronets}: $N_\text{s}^{vec}$= 16, $N_\text{out}^{vec}$=16, $\{L_\text{s}^{max}$=4, $N_\text{s}^{max}$=32, $N_\text{p}^{max}$=32, $N_\text{out}^{max}$=256, $N_\text{in}^{max}$=128$\}$. 
In Figure~\ref{fig:pq_area}, we use 16 bits as a baseline, and we vary the bitwidths of the distance calculators and product lookup portions of PQA to quantify the impact on FPGA resources utilization.
When quantizing the distance calculators, both the inputs and the prototypes share the same bitwidth.

Frequency remained relatively stable for each network when varying either the prototype or LUT bitwidth, according to Table \ref{tab:pq_vs_baseline}, and was often limited by the OpenCL \textit{shell} or board support package (BSP) that contains the infrastructure logic to connect PQA to PCIe and DDR4.
Frequency of DW$_\text{EMNIST}$ is notably slower due to the its larger area consumption.

Lower bitwidths for both distance calculation and product lookup significantly decrease area usage compared to the baseline as shown in Figure \ref{fig:pq_area}. 
The distance calculation has a steadily increasing linear trend for ALMs when varying the prototype bitwidth. 
The offset in the trend at 9 bits is attributed to the synthesis of the input buffer as BRAM instead of MLABs as the input size increases. 
For 2--3 bit wide inputs and prototypes, 0 DSPs are synthesized due to the low arithmetic complexity. 

When analyzing area trends in the product lookup operation, we account for both the lookup and accumulate kernels in order to capture any savings in the adder tree area when quantizing $LUT_{PQ}$ entries. 
The number of ALMs increases linearly due to both the increased accumulation size and the larger pipelined ``never-stall" load store unit (LSU). 
The increase in BRAM is mainly attributed to a larger $LUT_{PQ}$ at higher bitwidths.
Additionally, some BRAM is also utilized for storing the accumulator state as well but that does not increase with bitwidth as the accumulators are always fixed at 16 bits.

\tempremoved{
\begin{figure}[htbp]
\centerline{
\includegraphics[width=0.65\columnwidth]{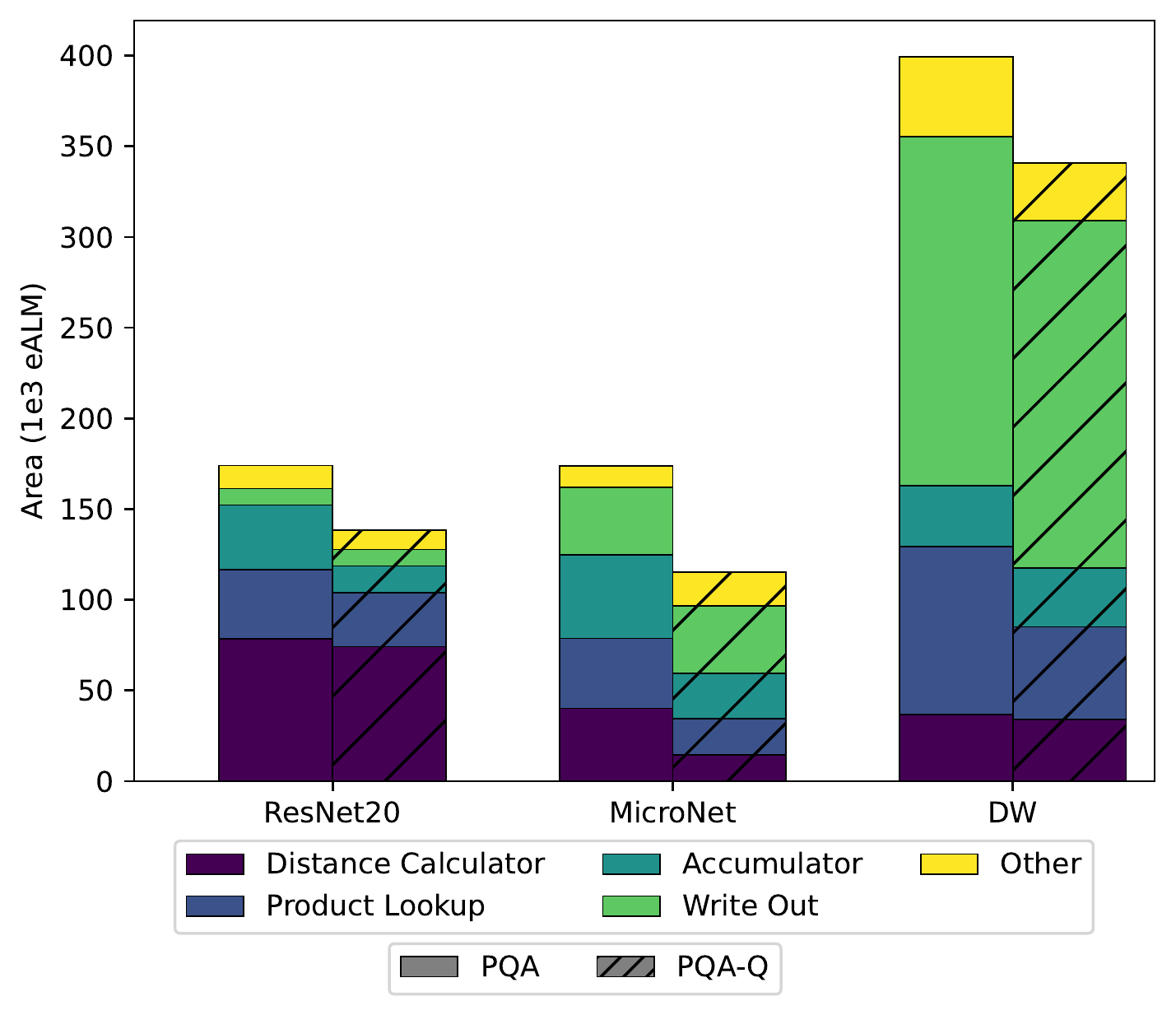}
}
\caption{Area breakdown of PQA vs PQA-Q for ResNet20$_\text{CIFAR10}$, MicroNet$_\text{KWS}$, and DW$_\text{EMNIST}$ with $<$1\% accuracy loss\pr{modified plot}}
\label{fig:area_breakdown}
\end{figure}
}

As Figure~\ref{fig:accuracy_plot} shows, lower bitwidths can be used without substantially impacting accuracy, especially with per-subspace quantization parameters. 
Per-subspace scale and offset results in the best accuracy at lower bitwidths while, at higher bitwidths, both per layer and even global scale and offset values yield comparable accuracy. 
By carefully adjusting the scale factors and clipping thresholds during per-subspace quantization, MicroNet$_{KWS}$ can use as little as 2 bits without any discernible effect on accuracy.
We use the combined results of Figures~\ref{fig:accuracy_plot} and \ref{fig:pq_area} to customize three PQA variants for our three PQ-DNNs: MicroNet$_\text{KWS}$, DW$_\text{EMNIST}$, and ResNet20$_\text{CIFAR10}$.
\tempremoved{
Figure~\ref{fig:area_breakdown} shows the relative areas and the area breakdowns of these customized versions of PQA compared to the non-quantized baselines with (8,16)~(bits
).
As Figure~\ref{fig:area_breakdown} shows, PQA quantization results in $\sim$25-50\% area savings compared to the baselines with minimal accuracy loss.
}

\subsection{PQA Acceleration Potential}

\begin{table}[t]\centering
\caption{Results of our PQ on multiple networks. Hardware latency is shown for DLA (baseline)\cite{dla_new} and PQA for two differerent external memory chips: DDR4 and HBM with 36 GB/s and 460 GB/s bandwidth respectively. PECAN-D\cite{pecan} is the current state of the art in PQ. PQA-Q is the quantized version of our accelerator with its parameters as bitwidths of prototypes and product tables respectively.}\label{tab:pq_vs_baseline}
\begin{tabular}{ccccccccccccc}\toprule
\multirow{2}{*}{Model} &\multirow{2}{*}{Setting} &\multirow{2}{*}{Ls} &\multirow{2}{*}{Np} &\multirow{2}{*}{Param.} &\multirow{2}{*}{\makecell{Acc \\ (\%)}} &\multirow{2}{*}{\makecell{Fmax \\ (MHz)}} &\multirow{2}{*}{\makecell{Area \\ {\small(keALMs)}}} &\multicolumn{2}{c}{\makecell{Latency \\ (us)}} &\multicolumn{2}{c}{\makecell{Performance/Area \\ (input/s/keALM)}} \\\cmidrule{9-10}\cmidrule{11-12}
& & & & & & & &DDR4 &HBM &DDR4 &HBM \\\toprule
\multirow{9}{*}{ResNet20$_\text{CIFAR10}$} &Baseline &-- &-- &269k &92.6 &485 &290 &37 &36 &94 &95 \\
\cmidrule{2-12}
&\multirow{2}{*}{PECAN-D} 
  &3 &64 &5.7M &87.9 &451 &217 &448 &276 &10 &17 \\
& &9 &64 &1.93M &85.0 &482 &227 &143 &86 &31 &51 \\
\cmidrule{2-12}

&\multirow{2}{*}{PQA} 
 &9 &16 &476k &84.4 &490 &225 &35 &24 &127 &185 \\

&  &9 &8 &239k &84.1 &471 &174 &28 &25 &207 &230 \\
\cmidrule{2-12}

&\multirow{2}{*}{PQA-Q (6,5)} 
  &9 &16 &476k &84.1 & 487 & 186 & 24 & 24 & 222 & 222 \\
& &9 &8 &239k &83.8 & 475 & 138 & 25 & 25 & \textbf{291} & 291 \\
\midrule

\multirow{6}{*}{MicroNet$_\text{KWS}$} &Baseline &-- &-- &61k &95.3 &485 &290 &4 &4 &885 &885 \\
\cmidrule{2-12}

&\multirow{2}{*}{PQA} 
  &4 &16 &212k &95.0 & 465 & 183 & 11 & 11 & 496 & 509 \\
& &8 &8 &63k &93.6 & 428 & 174 & 6 & 6 & 983 & 983 \\
\cmidrule{2-12}

&\multirow{2}{*}{PQA-Q (2,6)} 
   &4 &16 &212k &94.7 & 481 & 119 & 10 & 10 & 808 & 808 \\
&  &8 &8 &63k &93.3 & 473 & 118 & 5 & 5 & \textbf{1599} & 1599 \\
\midrule
\multirow{5}{*}{DW$_\text{EMNIST}$} &Baseline &-- &-- &1.1M &90.4 &485 &290 &47 &13 &74 &259 \\
\cmidrule{2-12}
&\multirow{2}{*}{PQA} &4 &12 &3.1M &87.6 & 288 & 746 & 231 & 51 & 6 & 26 \\
& &8 &8 &1.1M &85.8 & 287 & 399 & 82 & 27 & 31 & 93 \\
\cmidrule{2-12}
&\multirow{2}{*}{PQA-Q (5,5)} 
&4 &12 &3.1M &86.8 & 274 & 449 & 91 & 53 & 25 & 42 \\
&  &8 &8 &1.1M &85.0 & 306 & 341 & 34 & 25 & \textbf{87} & 116 \\
\bottomrule
\end{tabular}
\end{table}

In Table~\ref{tab:pq_vs_baseline}, we bring together our improved PQ training, efficient PQ-DNNs, and hardware PQA accelerator to assess the current state of product quantization on our three DNNs: ResNet20$_\text{CIFAR10}$, MicroNet$_\text{KWS}$, and DW$_\text{EMNIST}$.
We use DLA~\cite{aydonat2017opencl,dla_new} with conventional DNN execution as a baseline to which we can compare both the accuracy and performance of PQ-DNNs on PQA.
Furthermore, we are able to compare our ResNet20 PQ-DNN directly to prior work from PECAN~\cite{pecan}---the closest work on PQ in the literature.
To optimize PQA performance, we modified its vectorization parameters by setting $N_{out}^{vec}$ to 32 and by keeping other parameters at the minimum of 16 and the closest power of two of $L_s$ and $N_p$ as having a $L_s^{vec} > L_s$ or $N_p^{vec} > N_p$ is a waste of resources.
We additionally present results with the best quantization parameters discussed from Section~\ref{sec:quantization}, denoted with PQA-Q in Table \ref{tab:pq_vs_baseline}. 
Both designs have a higher Fmax and lower area with almost no accuracy drop, indicating the importance of reduced bitwidths to improve PQA performance and area. \revision{It's worth noting that the vectorization parameters can be further increased to reach even better performance in the cases where PQA-Q is compute-bound. For example, changing $N_s^{vec}$ from 16 to 32 in PQA-Q for $MicroNet_{KWS}$ in case of $L_s = 4$ and $N_p = 16$ improves the Performance/Area by around $1.5X$ showing that further vectorization parameters tuning can lead to even better gains in PQA.}

\noindent\textbf{Significant speedup is possible with PQ at lower accuracy.}
Focusing on ResNet20$_{\text{CIFAR10}}$, our largest DNN, we find that we are able to outperform both the baseline DLA, and PECAN considerably.
\ommission{This is because layer sizes are larger than the other two networks.}
Specifically, our best PQA-Q architectures achieved a 3.1$\times$ and 9.4$\times$ improvement in overall performance/area compared to the baseline DLA and PECAN alternatives.
While there is still a considerable gap to conventional DNNs in terms of accuracy, our improvements over PECAN-D comes at only 0.6\% accuracy degradation (at 4$\times$ performance/area boost) or up to 1.2\% degradation (with 9$\times$ performance/area boost). \revision{This is achieved due to the smaller $N_p$ and larger $L_s$ used and the efficiency of the optimized PQA-Q in performing the needed computations when compared to the conventional DLA that requires lots of multiplications.}
Our proposed efficient PQ-DNNs and PQA design therefore present a compelling accuracy-efficiency tradeoff for implementation on constrained edge devices.
Improvements can be seen in other DNNs as well where our best PQA-Q architectures achieve 81\% and 18\% improvement in overall performance/area for MicroNet$_{\text{KWS}}$ and DW$_{\text{EMNIST}}$ respectively.
Notably, the accuracy drop is lowest with MicroNet$_{\text{KWS}}$ with as little as 0.3\% degradation for some of our parameterizations.

\noindent\textbf{Lower bitwidth alleviates memory bandwidth bottlenecks.} 
As we have seen in our per-layer analysis, HBM favours PQA more than DLA because of the higher external memory bandwidth demands of PQ-DNNs.
However, when lower bitwidths were used with PQA, this is not the case anymore and the memory bandwidth bottleneck was alleviated as shown by the identical performance for both DDR4 and HBM for PQA-Q design points.

All considered, we were able to, for the first time, demonstrate a hardware speedup for PQ-DNNs, even when comparing against an optimized systolic array-based DLA.
Our results have shown that, while there is still an accuracy gap to conventional DNNs, our choice of PQ parameters, training improvements, and the \textit{accuracy corrector} can help decrease the accuracy degradation.
Our novel PQA architecture has helped to highlight that significant hardware speedup may be possible even if CPU and GPU architectures are not a good fit, especially when customizing the bitwidths for this new DNN computing paradigm as we have shown in our work.

\comment{
As expected, the faster HBM memory alleviates the memory bottleneck with PQ workloads and achieves higher speeds.
When normalizing for area, PQA achieves 40\% (DDR4) to 104\% (HBM) higher perf/area compared to DLA on ResNet20.
We demonstrate that it is possible to realize actual gains for PQ on hardware.
}

\section{Related Work}
\label{sec:related_work}

\textbf{Product Quantization.} Standard quantization approaches perform a scalar-to-scalar mapping while product quantization (PQ) operates with higher-dimensionality, mapping vectors to vectors~\cite{5432202,6678503}. 
This has made PQ a good fit for applications such as image retrieval~\cite{Yu_2018_ECCV,Klein_2019_CVPR, Jang_2020_CVPR} and compression~\cite{Stock2020And, image_PQ}. 
Different from those works is \cite{pmlr-v139-blalock21a}, where PQ is used to accelerate matrix-matrix multiplications in a two-step process: columns of the input are mapped to \textit{prototypes}, a set of learnable vectors; then these vectors are used to construct a look-up table of pre-computed dot products between prototypes and the layer weights. 
At inference time, the pre-computed values can be retrieved by mapping each input column to its closest prototype, trading numerical degradation for larger speedups in some cases. 
Follow up work extends PQ to an entire fully connected network and provides a preliminary analysis on the overheads of accelerating PQ~\cite{pq_all_you_need}. 
A more generalised implementation of PQ is PECAN~\cite{pecan}, which not only applies PQ to CNNs but also proposes a distance metric for input-to-prototype encoding that does not require multiplications. 
However, this method resulted in severe memory overheads since minimising accuracy degradation was prioritized. 
Unlike prior work, we perform a broader and systematic study of the impact of PQ settings in terms of memory, compute, and accuracy which we then use to inform our hardware accelerator design. Furthermore, we present the first hardware architecture to accelerate PQ-DNNs, and we demonstrate significant performance gains compared to prior work and to conventional DNNs.

\textbf{Hardware Acceleration of DNNs.}
Many custom hardware accelerators have been recently developed, especially for DNNs.
These accelerators outperform conventional CPUs and GPUs by leveraging DNN-specific properties in their hardware architectures~\cite{eie,eyeriss,adbutterfly}.
For example, Google's TPU~\cite{tpu} uses a 2D systolic array of multiply-accumulate units that can more directly and efficiently transfer data between compute units instead of expensive synchronization over GPU register files.
Another example is Groq's TSP~\cite{groq}, which utilizes spatial compute units to enable the construction of custom compute engines for each DNN layer.
Many prior works have used field-programmable gate-arrays (FPGAs) to build accelerators for deep learning, leveraging low precision, sparsity, a custom memory hierarchy, or novel dataflows~\cite{aydonat2017opencl,dla_new,hpipe}.

In the literature, other accelerators based on similar quantization techniques like vector quantization have been proposed~\cite{8702105}. However, this work is fundamentally different than our work. Their vector quantization algorithm is different from our product quantization algorithm. Specifically, we quantize activations on the fly and leave weights unquantized, while prior work\cite{8702105} quantizes weights and does not quantize activations. This significantly changes the hardware design and has implications on accuracy which are not quantified in the prior work unlike our work which focuses on finding product quantization parameters and optimizing the training process to achieve good accuracy. Prior work is also based on STD cell ASIC implementation, and area estimates are only based on post-synthesis results and without SRAM area included, as shown in the footnote in Table 1. However, the PQ accelerator (both in our work and in \cite{8702105}) is heavily-based on SRAM for product lookup. This makes the results in prior work much more difficult to compare with non-PQ accelerators. In contrast, our work is based on FPGAs and we are able to directly compare to a widely-used conventional deep learning accelerator to understand the true efficiency gap between the two approaches (PQA vs DLA).

To the best of our knowledge, none of the existing work has attempted to accelerate PQ using custom hardware\footnote{With the exception of a 1-page abstract at FCCM 2018~\cite{pq-cnn-poster}. However, without a description of the architecture.}, nor were there any studies of PQ efficiency using a custom accelerator.
Our work aims to fill this gap by presenting the first product quantization accelerator (PQA).

\vspace{-0.1cm}
\section{Conclusion} %
\label{sec:discussion}
\vspace{-0.1cm}

In this work we have identified several practical limitations that prevent PQ from being treated like other, more mature, optimizations techniques for DNN acceleration. 
As evidenced throughout our study, PQ requires a careful tuning to avoid incurring large overheads, especially in terms of memory and compute footprint. 
Unlike CPUs and GPUs, our custom hardware PQA has demonstrated that PQ can indeed accelerate entire DNNs but not as much as higher-level proxy metrics such as FLOPs might suggest. 
PQA is 3.1$\times$ more efficient than a conventional DLA in terms of performance/area on ResNet20.
To our knowledge, this is the first time a hardware speedup for PQ was demonstrated on any hardware platform.
Furthermore, by codesigning PQ parameters and hardware, our PQ ResNet20 is 4$\times$ better in performance per area than the most recent PQ work with just 0.6\% lower accuracy.
We also demonstrated the use of lower bitwidths through linear quantization on top of PQ to further decrease hardware area with low drop in accuracy.
In the future, we plan to investigate more robust and faster methods for PQ training, and explore more \textit{complementary} compression methods such as channel pruning.

\section{Limitations and Future Work}
\label{app:limitations}

In this work we made use of PQ to accelerate relatively small ML models for image classification and keyword spotting. Their size and complexity allowed us to consider a broad range of experiments, including an extensive per-layer analysis. An obvious extension of our work would be to apply PQ to much larger models.
Even though our evaluation focuses on relatively small networks and datasets, these networks are not over-parameterised for their respective tasks, meaning that achieving further compression is challenging without incurring accuracy degradation. 
This being said, compression is not the only objective of PQ. Our objective is to speedup inference through PQ without incurring a large increase in memory footprint due to the large  lookup tables. Currently, training PQ models requires much longer training times and memory utilization during the input-to-protoype encoding stage that require computing the distances of each input column of each subspace to each prototype. For example, training the MicroNet with PQ for SpeechCommands takes approximately 6 hours for one run, already making it challenging to produce most of the results in our paper. Not to mention, that PQ is highly sensitive to hyperparameter settings, which required us to run hundreds of hyperparameter search steps for each architecture-dataset pair.

The use of PQ is complementary to other forms of accelerating inference. For example, using quantization (as discussed in Section~\ref{sec:quantization}) could be used to reduce the computational footprint of the encoding stage as well as the overheads of storing and reading from LUT$_{\text{PQ}}$ \revision{and the external memory bandwidth limitations. This opens the door for exploring designs with higher vectorization parameters which may lead to higher speedups. In addition to that,} Structured pruning, in particular channel pruning~\cite{he2017channel, MLSYS2020_d2ddea18}, can be applied first to a model and then obtain its PQ representation. With channel pruning, the number of output channels is reduced and so the acceleration potential of PQ increases. This was shown analytically when analysing the trade-offs of PQ and empirically in our hardware layer-wise analysis.

\begin{acks}
\revision{This project is supported in part by Intel Corporation funding. We
would like to thank Deshanand Singh, Susanne Balle, Mahesh Iyer, Nilesh Jain, Aravind Dasu, Gregg
Baeckler, and Ilya Ganusov for insightful discussions and feedback. We would also like to thank the reviewers for their valuable comments and suggestions, which helped us improve the quality of the manuscript.}
\end{acks}

\bibliographystyle{ACM-Reference-Format}
\bibliography{sample-base}

\newpage
\appendix

\section{Model Architectures}
\label{sec:model_arch}
The per-layer study presented in the main paper made use of a CNN with 10 depth-wise separable layers. We designed this network, that we name $\text{DW}_{\text{EMNIST}}$, to be lightweight but deep enough so we could conduct the study considering a very large set of PQ-related parameters, \{$N_{\text{p}}$, $ L_{\text{s}}$, $\tau$\}, and training related hyperparamters (batch sizes, learning rates, etc). A detailed description of the network is presented in Table~\ref{tab:arch_EMNIST}. This network has a total of 1.05M paramters and requires 50.1M FLOPs when implemented as {\tt im2col}. In Section~\ref{sec:exp_evaluation} we analyse how a NAS-optimized MicroNet~\cite{MLSYS2021_a3c65c29}\footnote{open sourced in: \url{https://github.com/ARM-software/ML-zoo/tree/master/models/keyword_spotting/micronet_small/tflite_int8}} for keyword spotting performs under different PQ settings. We refer to this network as MicroNet-PQ and a per-layer breakdown of parameters and shapes is provided in Table~\ref{tab:arch_KWS}. This MicroNet baseline is an INT8 model designed to be run on microcontrollers. The last model considered in this work is a ResNet20~\cite{He_2016_CVPR} for CIFAR-10. This is the same model evaluated in PECAN~\cite{pecan}, and Table~\ref{tab:arch_ResNet20} provides a detailed view of this architecture.

\begin{table*}[t]
    \centering
    \caption{The $\text{DW}_{\text{EMNIST}}$ architecture used in the per-layer study designed for EMNIST image classification. Pointwise convolutions dominate the compute footprint of this model. Because of this, only point-wise convolutions are considered to be replaced with PQ -- for which we show the shapes of unrolled inputs and weights. This model has a total of 1.05M parameters and requires 50.1M FLOPs per $1\!\times\!28\!\times\!28$ input.}
    \scalebox{0.8}{
        \begin{tabular}{lcccccc}
            \toprule
            \textbf{Layer} & \textbf{Input Shape} & \textbf{Unrolled Inputs} & \textbf{Unrolled Weights} & \textbf{Parameters} & \textbf{FLOPs} & \textbf{FLOPs (\%)}\\
            \toprule
            Conv & [1, 1, 28, 28] & -- & -- & 640 & 1,003,520 & 2.00 \\
            \midrule
            DepthW-1 & [1, 64, 28, 28]  & --              & --               & 576        & 225,792    & 0.45 \\
            PointW-1 & [1, 64, 14, 14]  & [1, 64, 196]    & [96, 64]         & 6,144      & 2,408,448  & 4.81 \\
            \midrule
            DepthW-2 & [1, 96, 14, 14]  & --              & --               & 864        & 338,688    & 0.68 \\
            PointW-2 & [1, 96, 14, 14]  & [1, 96, 196]    & [120, 96]        & 11,520     & 4,515,840  & 9.01 \\
            \midrule
            DepthW-3 & [1, 120, 14, 14] & --              & --               & 1,080      & 423,360    & 0.85 \\
            PointW-3 & [1, 120, 14, 14] & [1, 120, 196]   & [150, 120]       & 18,000     & 7,056,000  & 14.08  \\
            \midrule
            DepthW-4 & [1, 150, 14, 14] & --              & --               & 1,350      & 132,300    & 0.26 \\
            PointW-4 & [1, 150, 7, 7]   & [1, 150, 49]    & [187, 150]       & 28,050     & 2,748,900  & 5.49 \\
            \midrule
            DepthW-5 & [1, 187, 7, 7]   & --              & --               & 1,683      & 164,934    & 0.33 \\
            PointW-5 & [1, 187, 7, 7]   & [1, 187, 49]    & [234, 187]       & 43,758     & 4,288,284  & 8.56 \\
            \midrule
            DepthW-6 & [1, 234, 7, 7]   & --              & --               & 2,106      & 206,388    & 0.41 \\
            PointW-6 & [1, 234, 7, 7]   & [1, 234, 49]    & [292, 234]       & 68,328     & 6,696,144  & 13.37 \\
            \midrule
            DepthW-7 & [1, 292, 7, 7]   & --              & --               & 2,628      & 84,096     & 0.17 \\
            PointW-7 & [1, 292, 4, 4]   & [1, 292, 16]    & [366, 292]       & 106,872    & 3,419,904  & 6.83 \\
            \midrule
            DepthW-8 & [1, 366, 4, 4]   & --              & --               & 3,294      & 105,408    & 0.21 \\
            PointW-8 & [1, 366, 4, 4]   & [1, 366, 16]    & [457, 366]       & 167,262    & 5,352,384  & 10.68 \\
            \midrule
            DepthW-9 & [1, 457, 4, 4]   & --              & --               & 4,113      & 131,616    & 0.26 \\
            PointW-9 & [1, 457, 4, 4]   & [1, 457, 16]    & [572, 457]       & 261,404    & 8,364,928  & 16.70 \\
            \midrule
            DepthW-10 & [1, 572, 4, 4]   & --              & --               & 5,148      & 41,184     & 0.08 \\
            PointW-10 & [1, 572, 2, 2]   & [1, 572, 4]     & [512, 572]       & 292,864    & 2,342,912  & 4.68 \\
            \midrule
            Linear & [1, 512]         & --              & --               & 24,111     & 48,222     & 0.10 \\
            \bottomrule

        \end{tabular}   
    }
    \label{tab:arch_EMNIST}
\end{table*}

\begin{table*}[t]
    \centering
    \caption{The MicroNet-PQ architecture designed for the SpeechCommands dataset. Pointwise convolutions dominate the compute footprint of this model, accounting for almost 75\% of the FLOPs. Because of this, only point-wise convolutions are considered to be replaced with PQ -- for which we show the shapes of unrolled inputs and weights. This model has a total of 60.6K parameters and requires 17.01M FLOPs per $1\!\times\!10\!\times\!49$ input. This model corresponds with MicroNet-KWS-S~\cite{MLSYS2021_a3c65c29}. We modified the number of channels of the second depth-wise block from 112 to 120 so it's divisible with common $\text{L}_\text{s}$ choices used throughout this work.}
    \scalebox{0.8}{
        \begin{tabular}{lcccccc}
            \toprule
            \textbf{Layer} & \textbf{Input Shape} & \textbf{Unrolled Inputs} & \textbf{Unrolled Weights} & \textbf{Parameters} & \textbf{FLOPs} & \textbf{FLOPs (\%)}\\
            \toprule
            Conv & [1, 1, 10, 49]  & --              & --               & 3,444      & 3,375,120  & 19.75 \\
            \midrule
            DepthW-1 & [1, 84, 10, 49] & --              & --               & 756        & 189,000    & 1.11 \\
            PointW-1 & [1, 84, 5, 25]  & [1, 84, 125]    & [120, 84]        & 10,080     & 2,520,000  & 14.75 \\
            \midrule
            DepthW-2 & [1, 120, 5, 25] & --              & --               & 1,080      & 270,000    & 1.58 \\
            PointW-2 & [1, 120, 5, 25] & [1, 120, 125]   & [84, 120]        & 10,080     & 2,520,000  & 14.75 \\
            \midrule
            DepthW-3 & [1, 84, 5, 25]  & --              & --               & 756        & 189,000    & 1.11 \\
            PointW-3 & [1, 84, 5, 25]  & [1, 84, 125]    & [84, 84]         & 7,056      & 1,764,000  & 10.32 \\
            \midrule
            DepthW-4 & [1, 84, 5, 25]  & --              & --               & 756        & 189,000    & 1.11 \\
            PointW-4 & [1, 84, 5, 25]  & [1, 84, 125]    & [84, 84]         & 7,056      & 1,764,000  & 10.32 \\
            \midrule
            DepthW-5 & [1, 84, 5, 25]  & --              & --               & 756        & 189,000    & 1.11 \\
            PointW-5 & [1, 84, 5, 25]  & [1, 84, 125]    & [196, 84]        & 16,464     & 4,116,000  & 24.08 \\
            \midrule
            Linear & [1, 196]        & --              & --               & 2,364      & 4,728      & 0.03 \\
            \bottomrule

        \end{tabular}   
    }
    \vspace{-0.2cm}
    \label{tab:arch_KWS}
\end{table*}

\begin{table*}[t]
    \centering
    \caption{The ResNet20 architecture for CIFAR-10. All layers in the network with the exception of the input convolution and output linear layer are replaced with their PQ counterparts in our evaluation. For these we show the shapes of unrolled inputs and weights. This model has a total of 268K parameters and requires 81.1M FLOPs per $3\!\times\!32\!\times\!32$ input.}
    \scalebox{0.8}{
        \begin{tabular}{lcccccc}
            \toprule
            \textbf{Layer} & \textbf{Input Shape} & \textbf{Unrolled Inputs} & \textbf{Unrolled Weights} & \textbf{Parameters} & \textbf{FLOPs} & \textbf{FLOPs (\%)}\\
            \toprule
            Conv & [1, 3, 32, 32]  & --              & --               & 432        & 884,736    & 1.09  \\
            \midrule
            Block1-Conv1 & [1, 16, 32, 32] & [1, 144, 1024]  & [16, 144]        & 2,304      & 4,718,592  & 5.82 \\
            Block1-Conv2 & [1, 16, 32, 32] & [1, 144, 1024]  & [16, 144]        & 2,304      & 4,718,592  & 5.82 \\
            Block1-Conv3 & [1, 16, 32, 32] & [1, 144, 1024]  & [16, 144]        & 2,304      & 4,718,592  & 5.82 \\
            Block1-Conv4 & [1, 16, 32, 32] & [1, 144, 1024]  & [16, 144]        & 2,304      & 4,718,592  & 5.82  \\
            Block1-Conv5 & [1, 16, 32, 32] & [1, 144, 1024]  & [16, 144]        & 2,304      & 4,718,592  & 5.82 \\
            Block1-Conv6 & [1, 16, 32, 32] & [1, 144, 1024]  & [16, 144]        & 2,304      & 4,718,592  & 5.82  \\
            \midrule
            Block2-Conv1 & [1, 16, 32, 32] & [1, 144, 256]   & [32, 144]        & 4,608      & 2,359,296  & 2.91 \\
            Block2-Conv2 & [1, 32, 16, 16] & [1, 288, 256]   & [32, 288]        & 9,216      & 4,718,592  & 5.82 \\
            Block2-Conv3 & [1, 32, 16, 16] & [1, 288, 256]   & [32, 288]        & 9,216      & 4,718,592  & 5.82 \\
            Block2-Conv4 & [1, 32, 16, 16] & [1, 288, 256]   & [32, 288]        & 9,216      & 4,718,592  & 5.82 \\
            Block2-Conv5 & [1, 32, 16, 16] & [1, 288, 256]   & [32, 288]        & 9,216      & 4,718,592  & 5.82 \\
            Block2-Conv6 & [1, 32, 16, 16] & [1, 288, 256]   & [32, 288]        & 9,216      & 4,718,592  & 5.82 \\
            \midrule
            Block3-Conv1 & [1, 32, 16, 16] & [1, 288, 64]    & [64, 288]        & 18,432     & 2,359,296  & 2.91\\
            Block3-Conv2 & [1, 64, 8, 8]   & [1, 576, 64]    & [64, 576]        & 36,864     & 4,718,592  & 5.82\\
            Block3-Conv3 & [1, 64, 8, 8]   & [1, 576, 64]    & [64, 576]        & 36,864     & 4,718,592  & 5.82\\
            Block3-Conv4 & [1, 64, 8, 8]   & [1, 576, 64]    & [64, 576]        & 36,864     & 4,718,592  & 5.82\\
            Block3-Conv5 & [1, 64, 8, 8]   & [1, 576, 64]    & [64, 576]        & 36,864     & 4,718,592  & 5.82\\
            Block3-Conv6 & [1, 64, 8, 8]   & [1, 576, 64]    & [64, 576]        & 36,864     & 4,718,592  & 5.82\\
            Linear & [1, 64]         & --              & --               & 650        & 1,300      & 0.00 \\
            \bottomrule

        \end{tabular}   
    }
    \vspace{-0.2cm}
    \label{tab:arch_ResNet20}
\end{table*}

\section{Extended PQ Implementation Trade-Offs}

\begin{figure*}[t]
    \centering
    \begin{subfigure}[b]{\textwidth}
    \centering
       \includegraphics[width=\linewidth]{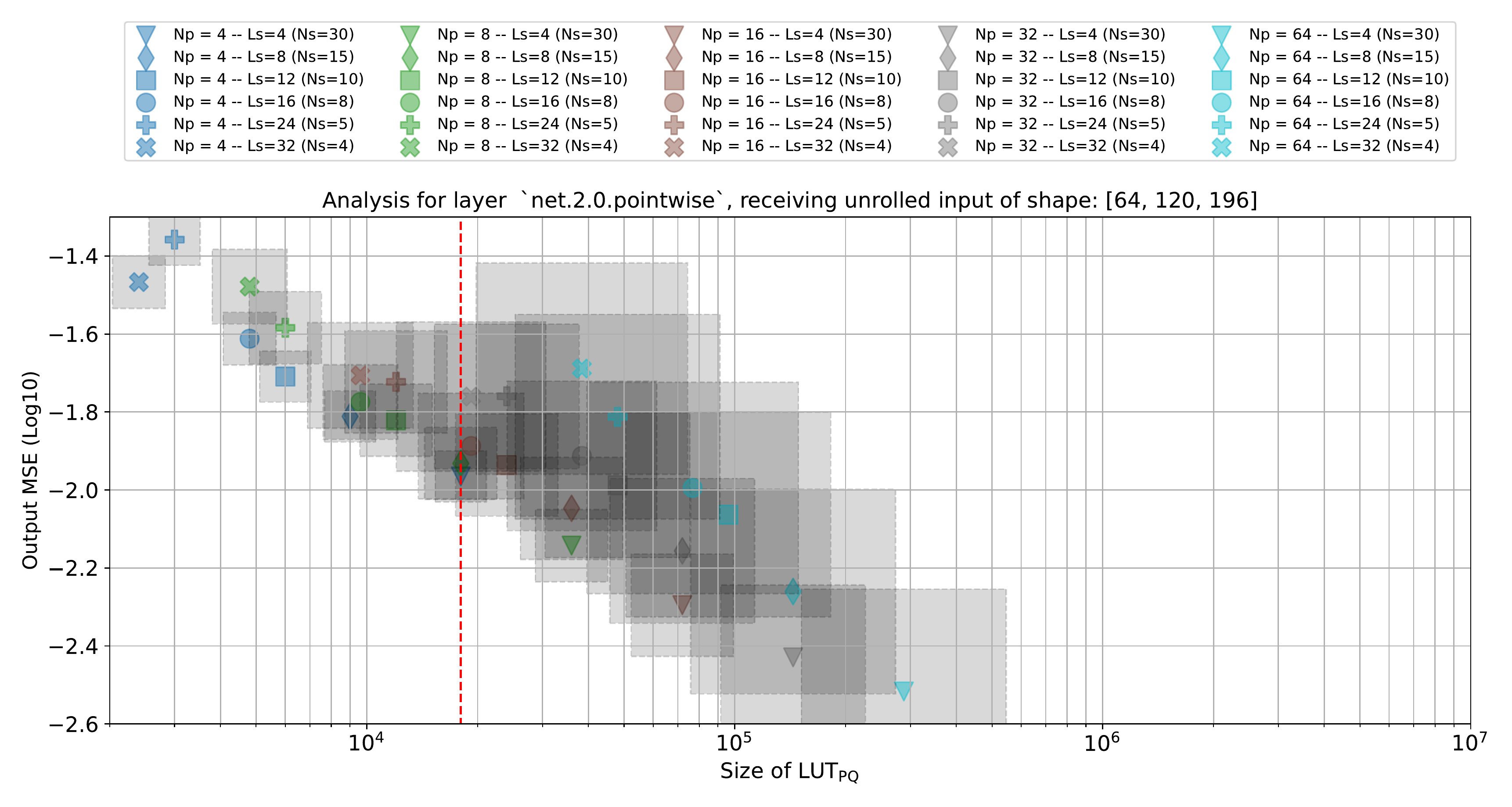}
    \end{subfigure}
    \vspace{-5mm}
    \begin{subfigure}[b]{\textwidth}
        \centering
       \includegraphics[width=\linewidth]{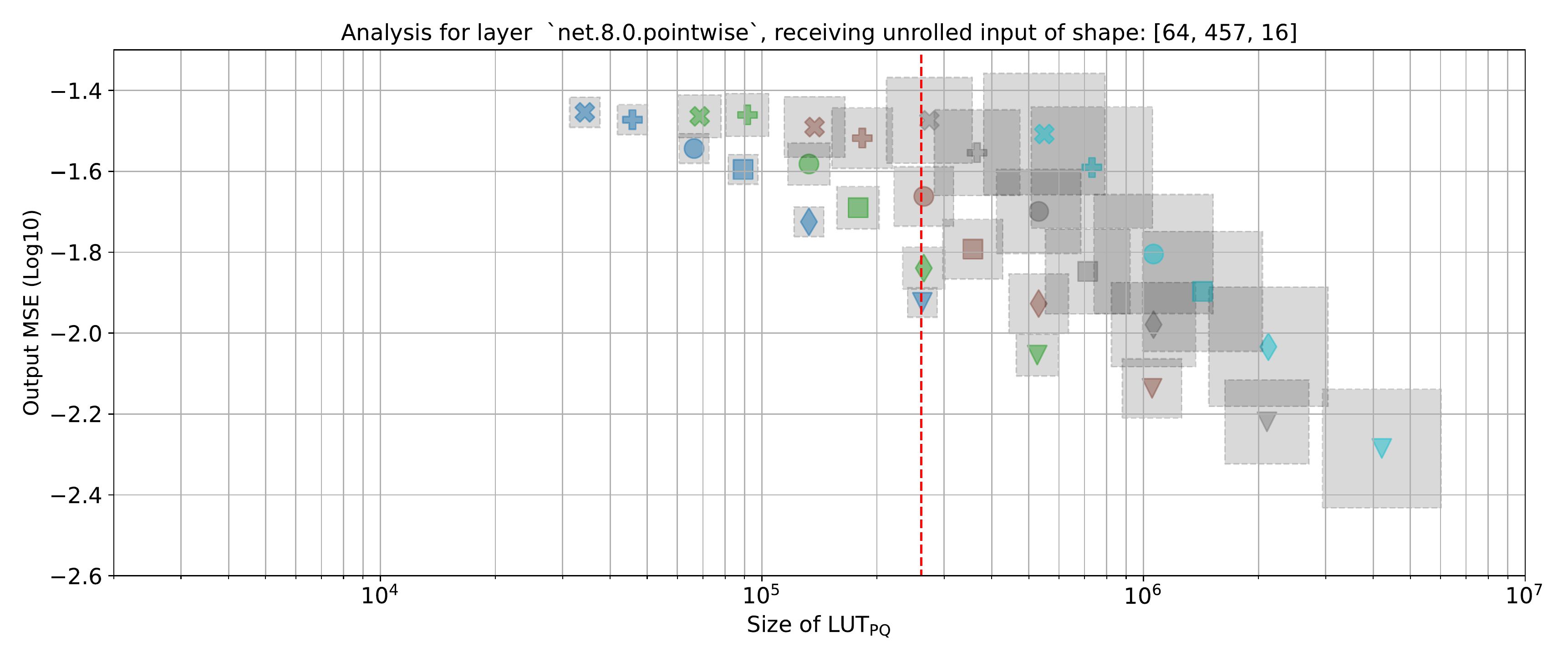}
    \end{subfigure}
    \vspace{-0.5mm}
    \caption{Analysis of the quality of the output generated by a PQ layer compared to the output of a non-PQ layer of a pre-trained model. The unit in the x-axis is number of parameters and we map to it the size of the resulting look up table, $\text{LUT}_{\text{PQ}}$, as well as the number of parameters that the same non-PQ layer has (this is shown as a vertical dashed red line). In addition to different memory footprints, each \{$N_{\text{p}}$, $L_{\text{s}}$\} pair also translates into different compute overheads when performing the input encoding. This is represented by the grey-shaded squares whose areas are proportional to $\text{FLOPS}^{\text{enc}}$.}
    \label{fig:overal_trend_extended}
\end{figure*}

In Fig.~\ref{fig:overal_trend_extended} we show an extended version of the results. Two main trends become evident in this visualisation. First, at deeper layers, where the spatial dimensions of the input tend to be smaller (\textit{i.e.} leading to fewer columns in the unrolled input) the impact on Output MSE of having more prototypes, higher $N_{\text{p}}$, is less pronounced. For example, in layer 2 and short prototypes $L_{\text{s}}\!=\!4$ having $N_{\text{p}}\!=\!64$ compared to $N_{\text{p}}\!=\!8$ offers an Output MSE $2.4\times$ smaller. In layer 8, this difference is reduced to $1.7\times$. The second trend has to do with the costs of performing the input-to-prototype indexing, or in other words, computing the distance of each input column with the bank of prototypes in a given subspace. Due to the larger spatial dimensions in early layers in the network, the encoding costs $\text{FLOPS}^{\text{enc}}$ are much larger for a constant \{$N_{\text{p}}$, $L_{\text{s}}$\} configuration than in deeper layers. For example for \{$N_{\text{p}}\!=\!64$, $L_{\text{s}}\!=\!8$\}, $\text{FLOPS}^{\text{enc}}$ is $3.2\times$ larger in layer 2 than layer 8.

\section{Memory and Compute footprint of PQ}
\label{app:Analyitical_footprints}

\begin{figure*}[h]
    \centering
    {
    \includegraphics[width=\linewidth]{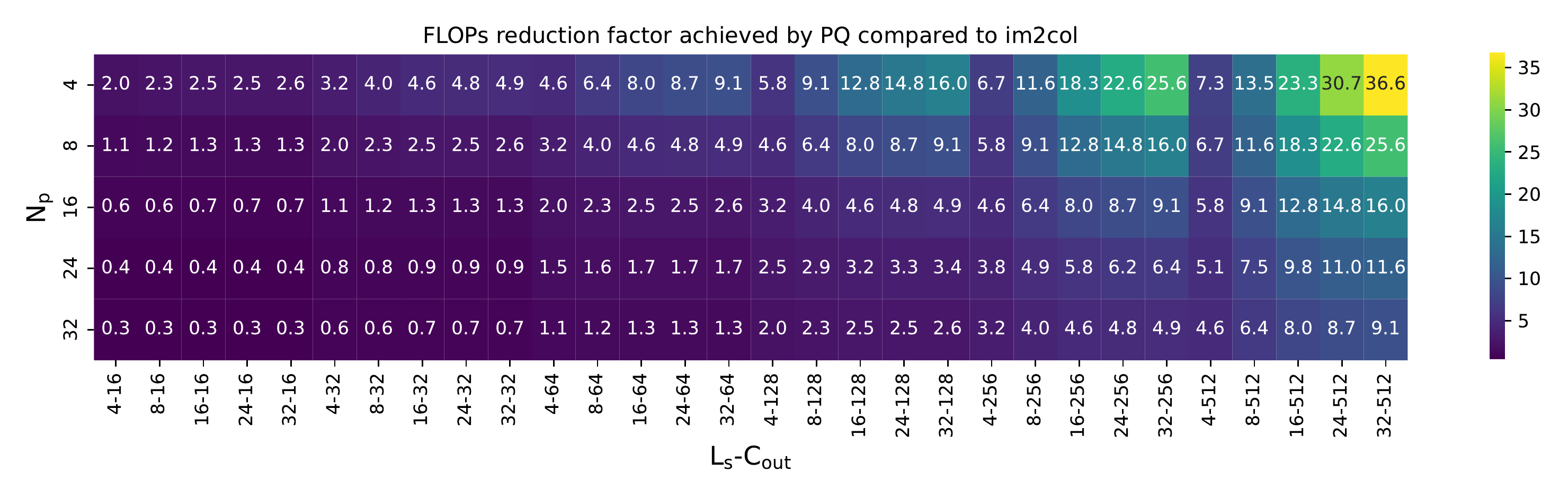}
    }
    \vspace{-5mm}
    \captionsetup{font=small,labelfont=bf}
    \caption{Reduction factor in the number of FLOPs when a layer with $C_{\text{out}}$ is implemented as a PQ layer with \{$N_{\text{p}}$, $L_{\text{s}}$\} parameterization. The speedup grows faster with $N_{\text{p}}$ than with $L_{\text{s}}$. For layers with very few output channels, there is no reduction in the number of FLOPs. Note this analysis assumes that looking-up the pre-computed dot product is cost-free.}
    \label{fig:flops_speedup}
\end{figure*}

We presented our study on the PQ trade-offs, we derived an expression for the speedup in terms of FLOPs achievable by PQ over {\tt im2col}-equivalent convolution in Section~\ref{sec:PQ_tradeoffs}. In Fig~\ref{fig:flops_speedup} we show how that expression maps to different speedups when varying $N_{\text{p}}$, $L_{\text{s}}$, and $C_{\text{out}}$. Having fewer prototypes (\textit{i.e.} a smaller $N_{\text{p}}$) has a larger impact than designing PQ with longer prototypes (\textit{i.e.} larger $L_{\text{s}}$), although this is also desirable. Another important observation is that PQ might not always be faster than {\tt im2col}. This is true for convolutional layers with few output channels ($C_{\text{out}}\! <\!64$) and a moderate number of prototypes per sub-space.

\section{Manhattan Distance}

\begin{figure*}[t]
    \centering
    \begin{subfigure}[b]{\textwidth}
    \centering
       \includegraphics[width=\linewidth]{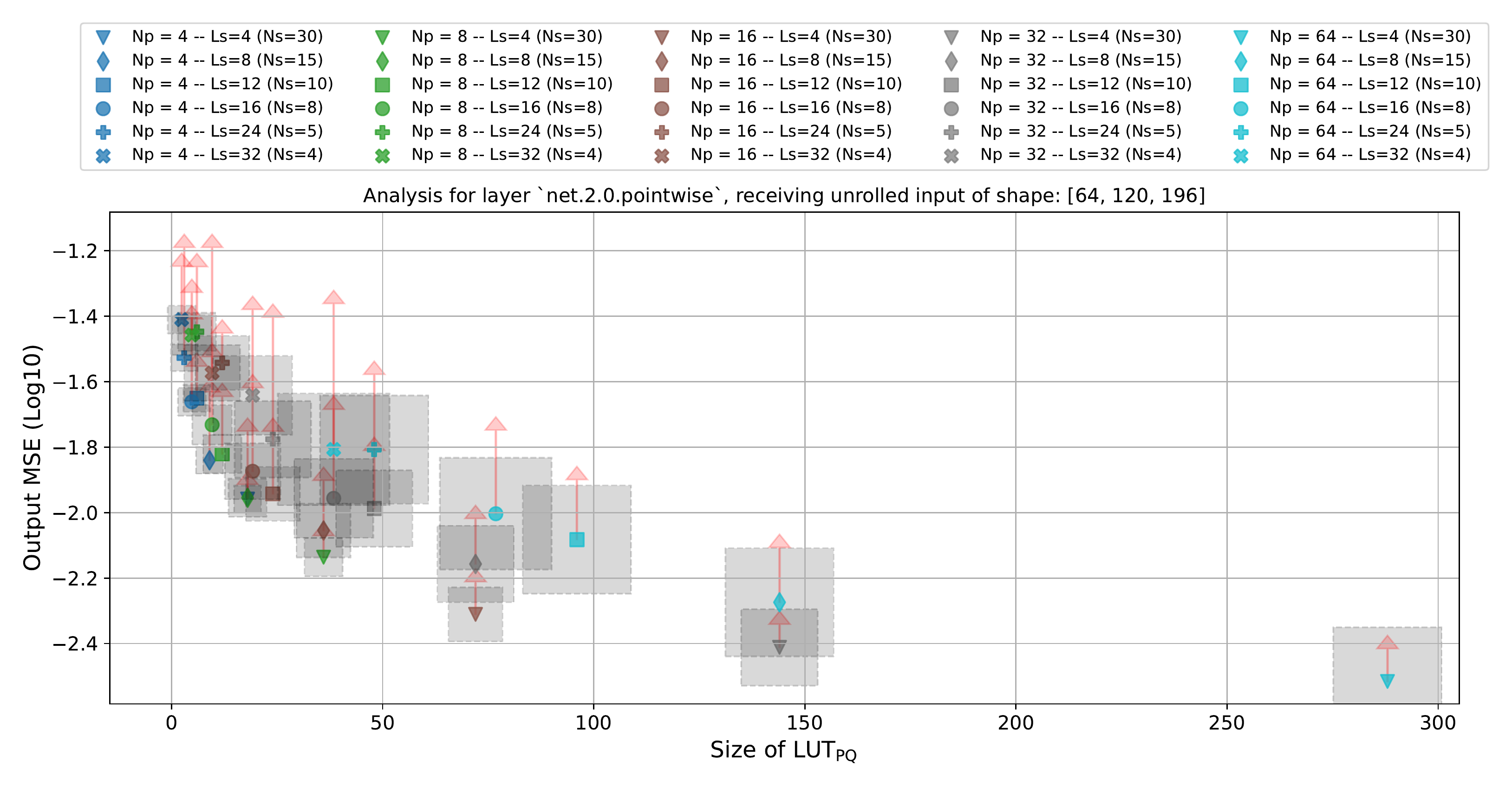}
    \end{subfigure}
    \vspace{-5mm}
    \begin{subfigure}[b]{\textwidth}
        \centering
       \includegraphics[width=\linewidth]{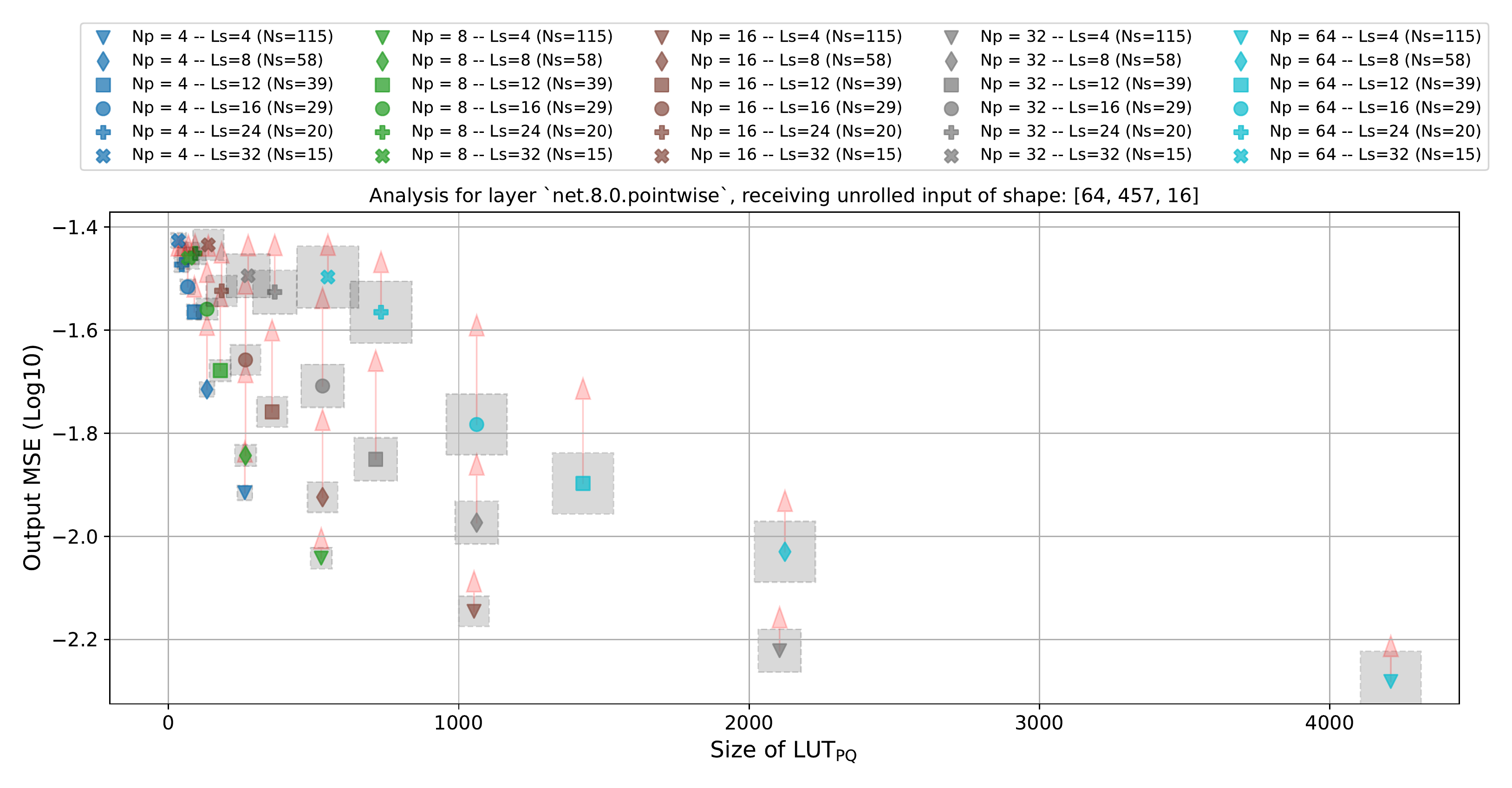}
    \end{subfigure}
    \vspace{-0.5mm}
    \captionsetup{font=small,labelfont=bf}
    \caption{When replacing Euclidean distance with the more lightweight Manhattan distance, the error introduced to the output (represented as a red arrow) is not negligible with longer prototypes. The size of LUT$_{\text{PQ}}$ is not affected by the choice of distance metric.}
    \label{fig:manhattan}
\end{figure*}

Figure~\ref{fig:manhattan} shows the impact of choosing Manhattan (L1) distance instead of Euclidean distance, which does not require multiplications. However, the impact on accuracy is not acceptable.

\ahm{Add back heatmaps after redrawing them correctly by adjusting the Number of Channels.}

\section{Accuracy and Hardware Performance}
\label{sec:accuracy_hardware_perf}

\begin{figure*}[t]
    \centering
    \begin{subfigure}[b]{0.6\textwidth}
    \centering
       \includegraphics[width=\linewidth, trim= 1cm 1cm 0 0]{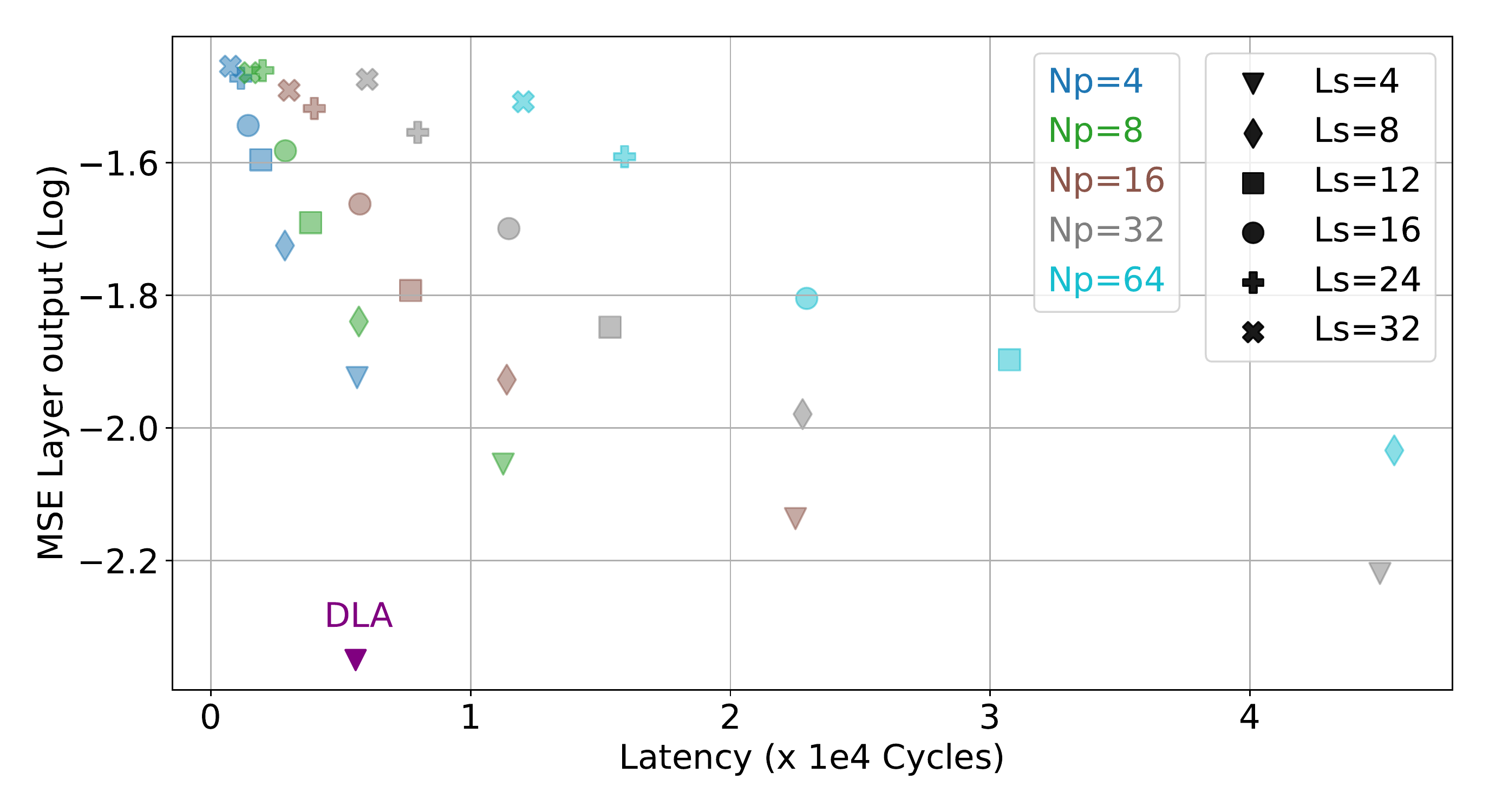}
    \end{subfigure}
    \vspace{-0.5mm}
    \captionsetup{font=small,labelfont=bf}
    \caption{\footnotesize MSE in a sample layer output vs number of cycles needed to process the input.}
    \label{fig:hw_mse_vs_cycle_count}
\end{figure*}

To link our hardware efficiency study with PQ accuracy, Fig~\ref{fig:hw_mse_vs_cycle_count} plots the the mean square error (MSE) of the output vs. the number of PQA cycles for different PQ parameters for PointW-9 layer described in Table~\ref{tab:arch_EMNIST}.
It can be seen that increasing $N_p$ increases the accuracy marginally while it has a clearer effect on the cycle count especially for larger $L_s$. 
It is worth noting that in this case, the efficiency of the hardware is limited by the external memory bandwidth so increasing $N_p$ increases the size of the lookup table leading to more memory bandwidth requirements.

\section{Comparison with other accelerators}

In the main text, a comparison with a conventional DLA~\cite{aydonat2017opencl} is presented. Here, comparison is extended to include another DLA~\cite{8891993} in Table~\ref{tab:pq_vs_dlas}. It’s clear that PQA with the correct configuration outperforms both which indicates that PQ running on a customized hardware can indeed have a benefit over conventional convolution running on a customized hardware. It’s worth mentioning that to be able to do the comparison with the information provided in the literature about the conventional accelerator~\cite{8891993}, some assumptions were done. We’re not taking into account the number of cycles taken by their engine to load data into the memory and we’re assuming it is always compute bound but we don’t make the same assumption in PQA (i.e. we count those cycles in PQA). One of the paper's optimizations is starting the calculation of the depthwise layer before the convolution layer before it is already done, they use the resultant outputs to kick off the computation in a pipeline manner. Given that there are no depthwise layers in ResNet20, we’re not overlapping any of the layers calculation with the other layers.

\begin{table*}[]
    \centering
        \captionsetup{font=small,labelfont=bf}
        \caption{Comparison between the proposed accelerator and conventional accelerators from the literature.}
    \scalebox{0.95}{
        \begin{tabular}{lc}
\hline
Accelerator   & Cycles per Image \\ \hline
PQA           & 11776            \\ \hline
\cite{aydonat2017opencl} & 17664            \\ \hline
\cite{8891993} & 19584            \\ \hline
\end{tabular}
    }
    \label{tab:pq_vs_dlas}
\end{table*}

\comment{
\moh{moved this here because we say this is done in the main text}

\begin{figure*}[t]
\centering
\begin{subfigure}[b]{0.65\textwidth}
\centering
   \includegraphics[width=\linewidth]{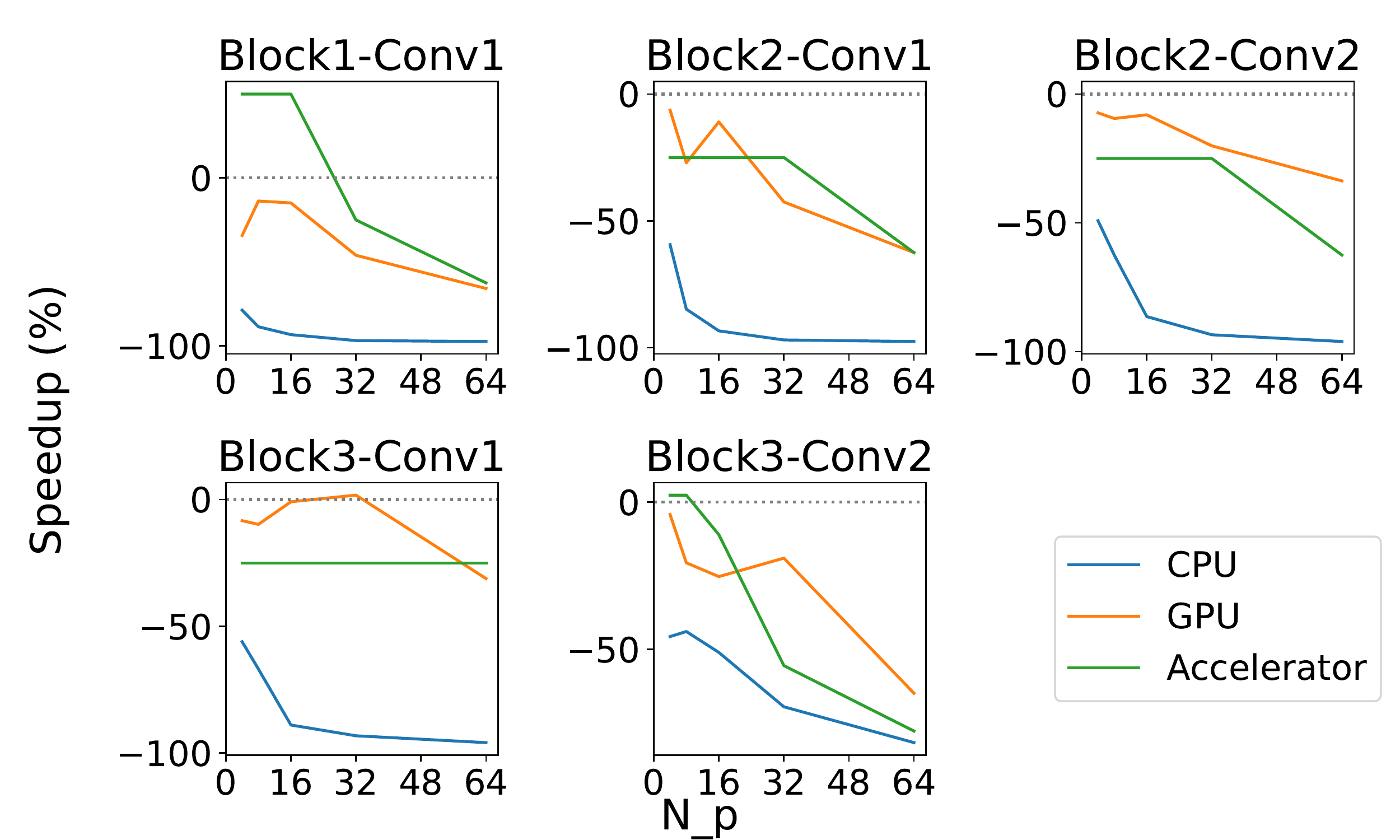}
\end{subfigure}
\vspace{-0.5mm}
\captionsetup{font=small,labelfont=bf}
\caption{\footnotesize Speedup of PQ vs conventional conv on different hardware when run on unique layers of PECAN Network. The value of $L_s$ here is kept at 16 and the speedup with different values of $N_p$ are shown.}
\label{fig:speedup_on_different_hardware_np}
\end{figure*}

Figure~\ref{fig:speedup_on_different_hardware_np} shows the same but when keeping value of $L_s$ at $16$ and changing the value of $N_p$. The same trends appear where CPU and GPU always show a slowdown while PQA shows sometimes speedup and sometimes slowdown.\pr{results aren't impressive. Shall we remove it?}
}

\end{document}